\documentclass[onecolumn,showpacs,preprintnumbers,amsmath,amssymb,a4paper]{revtex4} 

\usepackage{graphicx}    
\usepackage{dcolumn}     
\usepackage{bm}          
\usepackage[tight,TABTOPCAP,FIGTOPCAP]{subfigure}   

\graphicspath{{Figures/}}

\begin{document}




\title{Haze of surface random systems: An approximate analytic
  approach}

\author{Ingve Simonsen}
\email{Ingve.Simonsen@phys.ntnu.no}
\homepage{http://web.phys.ntnu.no/~ingves}
\affiliation{Department of Physics, Norwegian University of Science
  and Technology (NTNU), NO-7491, Trondheim, Norway}
\affiliation{NORDITA, Blegdamsvej 17, DK-2100 Copenhagen {\O}, Denmark}

\author{{\AA}ge Larsen}
\email{Age.Larsen@sintef.no}
\author{Erik Andreassen}
\email{Erik.Andreassen@sintef.no}
\affiliation{SINTEF Materials and Chemistry,
P.O. Box 124 Blindern, NO-0314 Oslo, Norway}

\author{Espen Ommundsen}
\email{Espen.Ommundsen@norner.no}
\affiliation{Norner Innovation AS, NO-3960 Stathelle, Norway}

\author{Katrin Nord-Varhaug}
\email{Katrin.Nord-Varhaug@recgroup.com}
\affiliation{REC Wafer Norway AS, Tormod Gjestlandsveg 31, NO-3908 Porsgrunn, Norway}

\date{\today}

\begin{abstract}
  Approximate analytic expressions for haze (and gloss) of Gaussian
  randomly rough surfaces for various types of correlation functions
  are derived within phase-perturbation theory. The approximations
  depend on the angle of incidence, polarization of the incident
  light, the surface roughness, $\sigma$, and the average of the power
  spectrum taken over a small angular interval about the specular
  direction.  In particular it is demonstrated that haze(gloss)
  increase(decrease) with $\sigma/\lambda$ as
  $\exp(-A(\sigma/\lambda)^2)$ and decreases(increase) with
  $a/\lambda$, where $a$ is the correlation length of the surface
  roughness, in a way that depends on the specific form of the
  correlation function being considered.  These approximations are
  compared to what can be obtained from a rigorous Monte Carlo
  simulation approach, and good agreement is found over large regions
  of parameter space. Some experimental results for the angular
  distribution of the transmitted light through polymer films,
  and their haze, are presented and compared to the analytic
  approximations derived in this paper. A satisfactory agreement is
  found.
  In the literature haze of blown polyethylene films has been related
  to surface roughness. Few authors have quantified the roughness and
  other have pointed to the difficulty in finding the correct
  roughness measure.

\end{abstract}

\maketitle


\section{Introduction}
\label{Sec:Intro}

Optical properties of polyethylene films have attracted considerable
attention due to the importance in applications such as packaging. 
Studies of haze and gloss of films have been carried out addressing
the effect of polymer structure~\cite{1x,2x,3x,4x}, rheological
properties~\cite{5x,6x}, additives~\cite{4x,7x,8x} and processing
conditions~\cite{2x,9x,10x,11x,12x,13x}.  

The notion of haze (``cloudiness'') is supposed to quantify the ratio
between the {\em diffusely} reflected or transmitted light to the {\em
  total} light reflected (reflectance) or transmitted (transmittance).
To this end, haze of a film is defined as the fraction of transmitted
light that deviates from the directly transmitted beam by more than
given amount ({\it e.g.}
$2.5^\circ$)~\cite{14x,HazeStandard1,HazeStandard2}. A similar
definition applies for reflection. For thin films, haze is recognized
to be caused mainly by scattering from surface irregularities, in
contrast to bulk randomness. Previously, only a few groups have
reported on the dependence of haze on surface
roughness~\cite{4x,12x,15x}. In a study of different polyethylene
materials, the bulk contribution to haze of $40 \mu m$ thick
films was found to vary in the range $10$--$30$\% of the total
value~\cite{4x}.  Two main mechanisms for surface roughness have been
identified~\cite{16x}: ({\it i}) flow-induced irregularities
originating from the die (extrusion haze), and ({\it ii}) protruding
crystalline structure such as lamellae, stacks of lamellae or
spherulites (crystallization haze).

Surface roughness is often characterized (in the engineering
literature) by the root-mean-square deviation from the mean surface
height, $\sigma$. Different roughness generating mechanisms, such as
die irregularities and inhomogeneous distribution of additives, will
influence the roughness at different length scales.  Large-scale
trends (compared to the wavelength) will not affect the diffuse light
scattering and must be eliminated from the analyzes.  Implicit
difficulties in deriving a relevant measure of surface roughness may
obscure its correlation with haze.  Various techniques have been
applied to characterize roughness.  In a number of recent studies
atomic force microscopy~(AFM) has been
applied~\cite{4x,10x,17x,18x,19x}.  Many authors, however, associate
only qualitative differences in roughness, as observed by {\it e.g.}
AFM, with haze values. Robust methods still have to be developed to
extract the relevant roughness measure based on a sufficiently high
statistics. A discussion of surfaces characterizing and relation to
haze is given in~\cite{4x}.

Sukhadia~{\it et al.}~\cite{20x} relate both crystallization haze and
extrusion haze to the elastic properties measured as the recoverable
shear strain of the polymer. For a low level of melt elasticity
orientations from the die relax quickly and crystalline aggregates
form on or close to the film surface. For highly elastic materials the
surface roughness was attributed to melt flow effects generated at the
die exit. Plotting haze versus the recoverable shear
strain~\cite{20x}, based on a large data set of blown and cast films,
results in a parabolic curve with lowest haze for materials of
intermediate level of elasticity.

Studies of electromagnetic wave scattering have a long history, and
the effect of surface roughness on the scattering has also been
studied for many years in
optics~\cite{Rayleigh,Beckmann,BassFuks,Ogilvy,Tsang-v1,Tsang-v2,Tsang-v3,Born,ThesisIngve,Ingve-condmat,Maradudin-Review,SA1,SA2},
and more recently, for x-ray
scattering~\cite{PhysRep1995,Sinha1988,deBoer,Andreev,Alex}.  Today it
is fair to say that the main features are rather well understood, at
least for one-dimensional roughness, where one recently has started to
address inverse (optical) problems~\cite{Designer1,Designer2}.  In the
case of two-dimensional roughness there are still open questions to be
answered, in particular for optical frequencies where the dielectric
contrast (and therefore the scattering) is the most pronounced.  This
paper concentrates on the range of optical frequencies, even if the
theoretical results derived herein hold for any frequency for which
the adopted approximations are expected to hold. This choice is made
due to the fact that the concept of haze (and gloss) mainly is used
within optics, to the best of our knowledge. 

Different approximations are available for scattering of
electromagnetic waves scattered by rough surfaces or bulk
inhomogeneities, depending {\it e.g.} on the typical range of
roughness/wavelength ratios ($\sigma/\lambda$), scattering angles, and
electrical
conductivity~\cite{14x,24x,25x,PhysRep1995,Sinha1988}. Vectorial or
scalar formulations are used, depending on whether or not polarization
is taken into account.  Vectorial formulations, such as the
Rayleigh-Rice perturbation theory, are considered to be more accurate
for smooth surfaces (typically $\sigma \ll \lambda$) with finite
conductivity, and large scattering angles. However, the scalar
Kirchhoff theory is claimed to be more accurate than Rayleigh-Rice for
rougher surfaces, and it is easier to handle
mathematically~\cite{24x}.

Some calculations of gloss {\it vs}. surface roughness parameters have
been published over the last years.  Alexander-Katz and
Barrera~\cite{Barrera} calculated the angular distribution of
reflected light, using the scalar Kirchhoff approximation. With a
Gaussian height distribution and various height correlation functions,
the gloss depended on two parameters. It increased with decreasing
$\sigma$ and increasing correlation length. The sensitivity of gloss
to correlation length depended on $\sigma$.  In particular, the
sensitivity was low for very low and high $\sigma/\lambda$ values.
Wang {\it et al.}~\cite{18x} performed similar calculations.  With
surface parameters (from AFM) and refractive indices as input, the
calculations agreed fairly well with gloss values in the range
$30$--$70$\%, measured on PE films.

Haze has been calculated as function of bulk
inhomogeneities~\cite{14x}, but theoretical literature on haze vs.
surface roughness is sparse, although there are related studies e.g.
on radio transmission~\cite{27x}.  Willmouth~\cite{14x} referred to
unpublished geometrical optics calculations relating film clarity to
surface roughness on a scale greater than the wavelength of light. In
most cases, however, the roughness is on a finer scale, and the
calculations must be based on physical optics.  The scalar Kirchhoff
approximation can, in principle, be applied to transmission
calculations~\cite{28x}, but this is more difficult than the
reflection case, due to contributions from two surfaces, possible bulk
effects, and a high probability of multiple scattering~\cite{24x}.

More recently Wang {\it et al.}~\cite{19x} presented a method for
calculating haze.  Since the (real) surfaces were spherulitic, the
combined scattering of the two surfaces could be modelled by Mie
scattering theory (valid for a single sphere of any size and
refractive index). Calculated haze values agreed with experimental
data for films of six different materials. Furthermore, the model
predicted a maximum in haze for a spherulite diameter of $800$~nm
($\lambda= 550$~nm), while the clarity decreased monotonously with
increasing spherulite diameter. This critical diameter, corresponding
to maximum haze, decreased with increasing refractive index, but it
was insensitive to the volume fraction of spheres, in the range
studied.  Although this model provides insight into the relationship
between surface roughness and haze, it has some limitations. In
particular, it can not account for non-spherical protrusions and
lateral correlation.

In this paper we study the effect of surface roughness on haze by
addressing both the height distribution as well as the height-height
lateral correlations. A motivation has been that problems in deriving
experimental measures of surface roughness have obscured the
correlation with haze~\cite{4x}. Our main goal is to establish the
relationship between surface structure --- both roughness
amplitude and surface lateral correlations --- and haze. An
analytical approximation to haze for Gaussian randomly rough surfaces
is derived which is compared with a rigorous Monte Carlo simulation
approach. Models are compared with experimental results on haze as
well as with the angular distribution of the transmitted light through
films.

This paper is organized as follows: The following section introduces
the scattering geometry to be discussed in this work as well as the
properties of the surface roughness. In Sec.~\ref{Sec:Scatt-Theory}
the scattering theory and notation to be used in the following
discussion is presented. The main results of this work starts in
Sec.~\ref{Sec:Haze} where the definition and analytic approximate
expressions for haze are presented. This approximation is compared to
rigorous computer simulations in Sec.~\ref{Sec:Results}. Finally the
conclusions that can be drawn from the present work are presented in
Sec.~\ref{Sec:Conclusions}


\section{The Scattering Geometry}
\label{Sec:Geometry}

The scattering geometry that we will consider in this study is
depicted in Fig.~\ref{fig:Geometry}. In the region $z > \zeta(x)$ it
consists of vacuum ($\varepsilon_{0}(\omega)=1$) and for $z <
\zeta(x)$ of a dielectric characterized by an isotropic,
frequency-dependent, dielectric function $\varepsilon_{1}(\omega)$. 
Here $\zeta(x)$ denotes the surface profile function. It is assumed
to be a single-valued function of $x$ that is differential as many
times as is necessary.  Furthermore, it constitutes a zero mean,
stationary, Gaussian random process that is defined by
\begin{subequations}
  \label{eq:surf-roughness}
\begin{eqnarray}
  \left< \zeta(x)\right> &=& 0, \\
  \left< \zeta(x)\zeta (x')\right> &=& \sigma^2 W(|x-x'|). 
\end{eqnarray}
\end{subequations}
Here $W(|x|)$ denotes the (normalized) auto-, or height-height
correlation function and will be specified later, $\sigma$ is the
root-mean-square of the surface roughness, and $\left<\cdot\right>$
denotes the average over an ensemble of realizations of the surface
roughness. For the later discussion one will also need the power
spectrum of the surface roughness, defined as the Fourier transform of
the correlation function, {\it i.e.} 
\begin{eqnarray}
  \label{eq:power-spec}
  g(|k|) &=& \int dx\, e^{-ikx}W(|x|). 
\end{eqnarray}
In this paper, we will mainly deal with correlation function of the
exponential type, $W(x) = \exp(-|x|/a)$, where $a$ is the so-called
correlation length. For such a correlation function the power spectrum
becomes
\begin{eqnarray}
  \label{eq:exp-corr}
   g(|k|) &=& \frac{2a}{1+k^2a^2}. 
\end{eqnarray}

The incident wave will be assumed to be either $p$- or $s$-polarized, as
indicated by the subscript $\nu$ on field quantities, and the plane of
incidence will be the $xz$-plane.  Furthermore, the angle of
incidence, scattering, and transmission, $\theta_0$, $\theta_s$, and
$\theta_t$ respectively, are measured positive according to the
convention indicated in Fig.~\ref{fig:Geometry}. 

In order to demonstrate that the assumption made above for the
statistics of the surface roughness is not unrealistic, we in
Fig.~\ref{fig:experimental-surf-data}(a) present an AFM measurement of
the surface topography of a polyethylene film surface.  The
corresponding height distribution and height-height correlation
function that can be obtained on the basis of such topography
measurements are can be found in
Figs.~\ref{fig:experimental-surf-data}(b) and (c). It is observed from
these figures that the measured surface roughness, to a good
approximation, is a Gaussian random process.  For this particular
example, an exponential correlation function is a reasonable, but not
perfect, choice for the correlation function.

\section{Scattering Theory}
\label{Sec:Scatt-Theory}

In this section, elements of the scattering theory that will be useful
for the discussion that will follow will be presented.  A more
complete and detailed presentation can be found {\it e.g.} in
Ref.~\cite{ThesisIngve}. 

\subsection{The reflection and transmission amplitudes}
\label{Sec:Amplitudes}

Due to the one-dimensional character of the surface topography,
$z=\zeta(x)$, a generic scalar field may be introduced that fully can
describe, together with Maxwell equations, the electromagnetic field. 
Such generic field is defined as
\begin{eqnarray}
  \phi_\nu(x,z,\omega) &=& \left\{
\begin{array}{ll}
  H_y(x,z|\omega), & \qquad \nu=p\\
  E_y(x,z|\omega), & \qquad \nu=s
\end{array}
\right. ,
\end{eqnarray}
where $\nu$ is a polarization index, and $H_y$ and $E_y$ are used to
denote the second component of the electric and magnetic fields. The
advantage of using the field variable $\phi_\nu(x,z,\omega)$ is that
the Maxwell (vector) equations satisfied by the electromagnetic field
are equivalent to the scalar wave equation for $\phi_\nu(x,z,\omega)$. 
To be able to use a scalar equation instead of vector equations
represents a great simplification of the problem. However, it should 
be noted that this simplification comes about because of $\zeta(x)$
being one-dimensional, and it doesn't hold true in general. 

For the scattering system considered in this paper, the field can be
written as
\begin{subequations}
  \label{eq:field-expansion}
\begin{eqnarray}
  \phi_\nu(x,z,\omega) &=& e^{ikx-i\alpha_0(k)z}
   + \int^{\infty}_{-\infty}\frac{dq}{2\pi}\, R_\nu(q|k)
  e^{iqx+i\alpha_0(q)z},
\end{eqnarray}
when $z>\max\zeta(x)$ and a plane incident wave is assumed,
and 
\begin{eqnarray}
  \phi_\nu(x,z,\omega) &=&  \int^{\infty}_{-\infty}\frac{dq}{2\pi}\, T_\nu(q|k)
  e^{iqx-i\alpha_1(q)z}
\end{eqnarray}
\end{subequations}
when $z<\min\zeta(z)$. In writing Eqs.~(\ref{eq:field-expansion}),
we have introduced $R_\nu(q|k)$ and $T_\nu(q|k)$ --- the reflection
and transmission amplitudes, the lateral momentum variable 
\begin{eqnarray}
  \label{eq:mom-angle}
  q &=& \sqrt{\varepsilon}\,\frac{\omega}{c}\sin\theta,
\end{eqnarray}
where $\varepsilon$ is the dielectric constant of the medium
considered, and $\theta$ denotes the angle of incidence, reflection or
transmission depending on context. 
Furthermore, in Eqs.~(\ref{eq:field-expansion}), we have also
defined
\begin{eqnarray}
      \alpha_m(q) &=&
      \left\{
        \begin{array}{cc}
          \sqrt{\varepsilon_m\frac{\omega^2}{c^2}-q^2}, &
                  \qquad |q|<\sqrt{\varepsilon_m}\frac{\omega}{c}\\
          i\sqrt{q^2-\varepsilon_m\frac{\omega^2}{c^2}}, &
                  \qquad |q|>\sqrt{\varepsilon_m}\frac{\omega}{c}
        \end{array}
     \right. ,
\end{eqnarray}
where $m=0$ corresponds to the medium above the rough surface, and
$m=1$ to the dielectric medium below.  Notice that when
$|q|\leq\sqrt{\varepsilon_m}(\omega/c)$ (non-radiative region), it
follows that $\alpha_m(q)=\sqrt{\varepsilon_m}(\omega/c)\cos\theta$.

\subsection{Mean Differential Reflection and Transmission Coefficients}
\label{Sec:MDRC}

The mean differential reflection and transmission coefficients,
abbreviated DRC and DTC, are two experimentally and theoretically
accessible quantities frequently used to study the angular
distribution of the reflected or transmitted light. We will here
denote them by $\left< \partial R_\nu/\partial\theta_s \right>$ and
$\left< \partial T_\nu/\partial\theta_t \right>$ respectively, where
subscript $\nu$, as before mentioned, is a polarization index.  The
mean DRC is defined as the fraction of the incident power that is
scattered by the rough surface into an angular interval $d\theta_s$
about the scattering angle $\theta_s$. The power (energy flux)
crossing a plane parallel to the $xy$-plane can be calculated from
\begin{eqnarray}
  \label{eq:Poynting}
  P &=&\int dx\!\int dy \;Re \left<{\bf S}_3\right>_t,  
\end{eqnarray}
where ${\bf S} =1/2 \, {\bf E}\times{\bf H}^*$ is the (complex) Poyntings
vector~\cite{Jackson} for the electromagnetic field $({\bf E}, {\bf H})$, and
$\left< \cdot \right>_t$ denotes a time-average. By using
Eqs.~(\ref{eq:field-expansion}) one finds that the incident power is
given by
\begin{subequations}  
  \begin{eqnarray}
    P_{inc} &=& \frac{L_1L_2}{2} \frac{c^2}{\omega} \alpha_0(k),
  \end{eqnarray}
  while the scattered power becomes
  \begin{eqnarray}
    P_{sc} &=& \varepsilon_0\frac{L_2}{2} \frac{c^2}{\omega} 
    \int^{\sqrt{\varepsilon_0}\frac{\omega}{c}}_{-\sqrt{\varepsilon_0}\frac{\omega}{c}}
    \!\frac{dq}{2\pi} \, \alpha_0(k) \left| R_\nu(q|k) \right|^2 
    \;=\;
    \int^{\frac{\pi}{2}}_{-\frac{\pi}{2}} d\theta_s \;
      p_{sc}(\theta_s).   
  \end{eqnarray}
  \end{subequations}  
  This latter relation implicitly defines the angular dependent
  scattered power $p_{sc}(\theta_s)$. Thus from the definition of the
  differential reflection coefficient it is realized that $\partial
  R_\nu/\partial\theta_s=p_{sc}(\theta_s)/P_{inc}$. However, since the
  surface is randomly rough, it is the {\it mean} differential
  reflection coefficient that should be of interest. Such a quantity
  will be given by
  \begin{eqnarray}
    \label{eq:mean-DRC}
    \left<\frac{\partial R_\nu}{\partial\theta_s}\right>
    &=& \left< \frac{p_{sc}(\theta_s)}{P_{inc}}\right> 
      \;=\; \frac{\sqrt{\varepsilon_0}}{L_1}
          \frac{\omega}{2\pi c}\frac{\cos^2\theta_s}{\cos\theta_0}
          \left<\left| R_\nu(q|k) \right|^2\right>,
  \end{eqnarray}
  where we have used $\left<\cdots\right>$ to denote the average over
  surface realizations.  In the same way, the mean differential
  transmission coefficient may be defined as $\left<\partial
    T_\nu/\partial\theta_t\right>=\left<p_{tr}(\theta_t)/P_{inc}\right>$,
  where $p_{tr}(\theta_t)$ denotes the power transmitted into an
  angular interval about the angle of transmission $\theta_t$. Instead
  of giving the expression for this quantity explicitly, we will instead for
  later convenience, introduce a generic notation for both the mean
  differential reflection and transmission coefficient. This generic
  quantity will be denoted by $\left<\partial U/\partial\theta\right>$
  so that
\begin{subequations}
  \begin{eqnarray}
  \label{eq:generic-DRC-DTC}
   \left<\frac{\partial U}{\partial\theta}\right>
     &=& \left\{
       \begin{array}{ccl}
          \left<\frac{\partial R_\nu}{\partial\theta_s}\right>, &
          \qquad & \mbox{in reflection} ,\\
          \left<\frac{\partial T_\nu}{\partial\theta_t}\right>, &
          \qquad & \mbox{in transmission}
        \end{array}
        \right. ,
\end{eqnarray}
where $\theta$ stands for $\theta_s$ and $\theta_t$ in reflection and
transmission, respectively. In general it may be written in the
following form
  \begin{eqnarray}
    \label{eq:mean-DRC-DTC}
        \left<\frac{\partial U}{\partial\theta}\right>
      &=& \frac{1}{L_1} \frac{\varepsilon_m}{\sqrt{\varepsilon_0}}
          \frac{\omega}{2\pi c}\frac{\cos^2\theta}{\cos\theta_0}
          \left<\left| U(q|k)\right|^2\right>. 
  \end{eqnarray}
\end{subequations}
In Eq.~(\ref{eq:mean-DRC-DTC}) $U(q|k)$ denotes the reflection or
transmission amplitudes depending on context, and $m$ takes on the
values $m=0$ in reflection and $m=1$ in transmission.

In theoretical studies it is customary to separate $\left<\partial
  U/\partial\theta\right>$ into two terms --- one coherent and one
incoherent term. That this is possible can be realized from the
following (trivial) rewriting
\begin{eqnarray}
  \label{eq:coh-incoh}
 \left<\left| U(q|k) \right|^2\right> =
   \left|\left<   U(q|k) \right>\right|^2 +
   \left[ \left<\left| U(q|k) \right|^2\right> -\left|\left<
   U(q|k) \right>\right|^2 \right]. 
\end{eqnarray}
If this expression is substituted back into
Eq.~(\ref{eq:mean-DRC-DTC}), the first term will give rise to the {\em
  coherent} or specular contribution to $\left<\partial
  U/\partial\theta\right>$, while the last term, within the square
brackets, will result in the {\em incohrent} or diffuse contribution
to the same quantity. We will use subscripts {\it coh} and {\it
  incoh}, respectively, to indicate these whenever needed. This
identification follows from observing that in the average
$\left<U(q|k)\right>$ one only gets contributions from those portions
of $U(q|k)$ that are {\it in phase} from one surface realization to
another.




From Eq.(\ref{eq:mean-DRC-DTC}) a quantity that defines the fraction
of the incident energy that is either reflected or transmitted can be
defined as
\begin{eqnarray}
  \label{eq:ref}
  {\cal U} 
      &=&  \int^{\pi/2}_{-\pi/2}\,d\theta \; 
         \left<\frac{\partial U}{\partial\theta}\right>. 
\end{eqnarray}
Notice, that when no absorption takes place in neither media involved,
{\it i.e.} $Im\, \varepsilon_m=0$, one should have that the sum of
this quantity in reflection and transmission should add up to one,
${\cal U}_{s}+ {\cal U}_{t}=1$. This is a direct consequence of energy
conservation. In practice, however, there will always be some
absorption, but for many dielectric media at optical frequencies it is
a reasonable approximation to neglect it.

\section{Haze}
\label{Sec:Haze}

In the optical industry two quantities --- {\em haze} and {\em gloss}
--- are often used to quantify the visual appearance of
materials~\cite{Gloss}.  Gloss, crudely speaking, is related to the
amount of light being reflected (or transmitted) into angles {\em
  around} the specular direction. This quantity has previously been
studied experimentally~\cite{Mendez} for transparent plastic
materials, and recently also studied theoretically~\cite{Barrera}. In
this latter study, the authors investigated how gloss depends on the
level of roughness and surface correlations. It was found that the
incoherent (diffuse) contribution to the scattered light could
contribute significantly to the gloss.  Haze, on the other hand,
measures the fraction of reflected (transmitted) light that is
reflected (transmitted) {\em away} from the specular
direction~\cite{HazeStandard1,HazeStandard2}.  In the former case one
talks of {\em haze in reflection} and in the latter of {\em haze in
  transmission}. Notice, that haze can almost be considered as a
complementary quantity to gloss.  If the haze of a transparent plastic
film, say, is large, then an object viewed through the film will look
unsharp or blurry. It is this kind of visual effect that the haze
value is supposed to quantify.

Even though gloss is important in many applications, we will in the
present study concentrate on haze since this quantity has not been
studied that extensively in the literature by theoretically means.

\subsection{Definition of haze}

For the purpose of this theoretical study, we will focus on a
semi-infinite medium, instead of a film geometry. The reason for this
choice is purely practical.  Let us start by assuming that the
incident light is impinged onto the planar mean surface at an angle
$\theta_0$ measured counter clock-vise from the normal to the mean
surface (cf.~Fig.~\ref{fig:Geometry}). In case of no surface roughness
the light will be scattered (or transmitted) in accordance with
Snell's law.  Hence all the energy will be propagate in the directions
defined by the angle
\begin{eqnarray}
  \label{eq:specular-direction}
  \Theta &=& \arcsin \left( 
              \frac{\sqrt{\varepsilon_m}}{\sqrt{\varepsilon_0}} 
              \sin\theta_0
              \right), 
\end{eqnarray}
where $m=0$ should be used in reflection (for which
$\Theta=\theta_0$), and $m=1$ in transmission. 

Formally, haze, ${\cal H}(\theta_0)$, is
defined~\cite{HazeStandard1,HazeStandard2,Billmeyer} as
the {\em fraction} of the reflected (transmitted) light that is
reflected (transmitted) into angles lying {\em outside} the angular
interval $(\theta_-,\theta_+)$ with
\begin{eqnarray}
  \label{eq:theta-ref}
  \theta_\pm=\Theta\pm\Delta\theta,
\end{eqnarray}
and where $\Delta\theta$ is an angular interval to be defined. In
commercially available
haze-meters~\cite{HazeStandard1,HazeStandard2,Billmeyer}, one for this
angular interval uses the value $\Delta\theta =
2.5^\circ$\footnote{The actual angular interval depends on the type of
  haze one wants to measure, {\it e.g.} narrow- or wide-angle haze,
  and it depends on the actual standard one desires to comply with.
  For instance, one should notice that the angular interval may be
  different for reflection and transmission.},
and for the present study this
value will be adopted. Notice that the angles $\theta_\pm$ can be
related to the lateral momentum variable, $q_\pm$, in accordance with
Eq.~(\ref{eq:mom-angle}). Furthermore, the angular interval
$\Delta\theta$ is related to a corresponding  momentum interval in
the following way
\begin{eqnarray}
  \label{eq:delta-mom}
  \Delta q &=& \frac{1}{2}(q_+-q_-)
      \;=\;
        \sqrt{\varepsilon_m}\,\frac{\omega}{c} 
         \cos\Theta \sin\Delta\theta . 
\end{eqnarray}

Haze, as defined above, can readily be related to the mean
differential reflection or transmission coefficients $\left<\partial
  U/\partial\theta\right>$ (cf. Sec.~\ref{Sec:MDRC}). In terms of
these quantities, haze, that in general will depend on the angle of
incidence $\theta_0$, can be written in the form
\begin{subequations}
  \label{eq:haze}
\begin{eqnarray}
  \label{eq:haze-A}
{\cal H}(\theta_0) 
   &=& I\left(-\frac{\pi}{2},\theta_-\right) +
        I\left(\theta_+,\frac{\pi}{2}\right) 
                         \\
        \label{eq:haze-B}
        &=&
         1 - I(\theta_-,\theta_+),
\end{eqnarray}
where one has introduced 
\begin{eqnarray}
  \label{eq:I-int}
  I(\theta_a,\theta_b)   &=&  
    \frac{1}{{\cal U}} \;
           \int^{\theta_b}_{\theta_a}d\theta \, 
           \left<\frac{\partial U}{\partial\theta}\right>,
 \end{eqnarray}
\end{subequations}
and where ${\cal U}$ has been defined earlier in Eq.~(\ref{eq:ref}). 
In the transition from Eqs.~(\ref{eq:haze-A}) to (\ref{eq:haze-B}) it
has been used that $I\left(-\pi/2,\pi/2 \right)=1$ by definition.  The
reason for the presence of the factor $1/{\cal U}$ in
Eq.~(\ref{eq:I-int}) is that haze is defined in terms of the fraction
of the reflected or transmitted light, while $\left<\partial
  U/\partial\theta\right>$ is defined as the fraction of the incident
power. Hence, this factor is present to ensure that the defining
expression of haze has been given the correct normalization. From the
definition, Eqs.~(\ref{eq:haze}), it follows that haze is a
dimensionless number between zero and one~\footnote{Haze is often also
  indicated by using a ``percentage notation'' so that a haze of $1$
  corresponds to a haze of $100\%$.}.  Furthermore, notice that an
ideal scattering system, {\it i.e.}, one with no surface or bulk
randomness, will correspond to a haze value of ${\cal H}(\theta_0)=0$
for all angles of incidence, since all light will be reflected or
transmitted into the specular direction~\footnote{One should, however,
  be aware that this is only fully true if a plane incident wave is
  used. On the other hand, if an incident beam of finite width is
  applied, one can, at least in principle, get a non-vanishing, but
  small, haze value even for an ideal scattering system.}.  However,
as the randomness of the scattering system is increased, and therefore
the reflected or transmitted intensities as a consequence become more
and more diffuse, the corresponding haze value will increase. To reach
a haze of one (${\cal H}=1$) is, however, rather unlikely for random
systems encountered in practical situations since it will require a
vanishing intensity over the angular interval $(\theta_- ,\theta_+)$. 
Surface random systems with such a property can, however, be
artificially manufactured~\cite{DesginerSurfaces}.  From a practical
point of view, a more likely scenario for a strongly random system is
probably that of a Lambertian diffuser~\cite{LambertianDiff}, {\it
  i.e.}, a scattering system giving raise to $\left<\partial
  U/\partial\theta \right>\propto\cos\theta$ independent of angle of
incidence. For such a scenario, haze at normal incidence will be
${\cal H}(0)=1-\sin\Delta\theta\simeq 0.956<1$ according to the
definition~(\ref{eq:haze}). For naturally occurring surfaces this is
probably a more realistic upper limit of haze.

\subsection{A naive approximation to haze}

In many situations encountered in practical applications of the
concept of haze, it is illuminative to have available an approximate
expression to the formal definition~(\ref{eq:haze}).  Often the
dependence on various parameters on that haze will depend, can be made
more apparent via approximate expressions. 

Before considering an analytic approximation to haze (see next
subsection) we will present some approximations that are based on the
concept of coherent and incoherent scattering. Such picture is often
useful to bear in mind when working with the haze (and gloss) concept.

If the scattering system is not too diffuse, the main contribution to
the $I$-integral present in Eq.~(\ref{eq:haze}b) will come from the
coherent component of the scattered or transmitted light. Furthermore,
since the coherent component is non-zero outside the angular interval
from $\theta_-$ to $\theta_+$, one arrives at the following
approximation to haze
\begin{subequations}
  \label{eq:approx}
\begin{eqnarray}
  \label{eq:approx-A}
 {\cal H}(\theta_0)
  &\simeq&  1-  \frac{1}{{\cal U}} \;
          \int^{\pi/2}_{-\pi/2}d\theta \, 
            \left<\frac{\partial U}{\partial\theta}\right>_{coh},
\end{eqnarray}
based exclusively on the coherent component of the scattered or
transmitted field. Thus, it is natural to refer to
Eq.~(\ref{eq:approx-A}) as a {\em coherent approximation to haze}. 
From a mathematical point of view, one may rewrite
Eq.~(\ref{eq:approx-A}) by noting that the last term is just one minus
the corresponding incoherent component. By recalling
Eqs.~(\ref{eq:coh-incoh}) and (\ref{eq:ref}) one, therefore, arrives
at the following equivalent {\em incoherent approximation to haze}
\begin{eqnarray}
  \label{eq:approx-B}
 {\cal H}(\theta_0)  
  &\simeq&  \frac{1}{{\cal U}} \;
          \int^{\pi/2}_{-\pi/2} d\theta \, 
            \left<\frac{\partial U}{\partial\theta}\right>_{incoh}. 
\end{eqnarray}
\end{subequations}

Later it will be shown that the coherent approximation,
Eq.~(\ref{eq:approx-A}), only takes into account the surface
roughness, and {\em not} its correlation. The incoherent
approximation, Eq.~(\ref{eq:approx-B}), however, will be shown to {\em
  also} depend on the surface correlation. Even though the coherent
and incoherent approximations to haze are equivalent from a purely
mathematical point of view, they give rise to different physical
interpretations. Since Eq.~(\ref{eq:approx-B}) includes the dependence
on the surface correlation,as well as surface roughness, we will
prefer this approximation over that of Eq.~(\ref{eq:approx-A}). 

An approximation is not worth much without information about its range
of validity.  In order to get an idea of the accuracy of the coherent
and incoherent approximations, the Lambertian diffuser will be
considered once more. Such a diffuser represents in many ways a worst
case scenario since there is no coherent component at all in this
particular case.  Due to the vanishing coherent component, one
therefore, within the approximations of Eqs.~(\ref{eq:approx}), has
that ${\cal H}\simeq 1$.  However, above one found by using the formal
definition of haze, Eqs.~(\ref{eq:haze}), that ${\cal H}\simeq 0.956$.
Hence, the error obtained by calculating the haze from the approximate
expressions~(\ref{eq:approx}) is of the order of $4.5\%$ for this
highly diffusive case. For less diffusive surfaces the error is
expected to be less.

\subsection{An analytic approach to haze}

In this subsection, an analytic expressions for haze will be derived.
This will be achieved by applying the so-called phase perturbation
theory~\cite{Shen,Rosa}.  This approximation is reviewed in
Appendix~\ref{App}, and it can, at least in reflection, be viewed as
an extension (or correction) to the more well-known (and used)
Kirchhoff approximation~\cite{Beckmann,BassFuks,Ogilvy,Tsang-v1}.  In
particular, phase-perturbation theoretical results reduce in the limit
of large correlation length for the surface roughness to those that
can be obtained from Kirchhoff theory.  In addition,
phase-perturbation theory naturally can handle transmission problems
in an analytic fashion while such a generalization is not straight
forward for the Kirchhoff approximation.

Within phase-perturbation theory, the mean DRC or DTC, $\left<\partial
  U/\partial\theta\right>$, can be written in the following form (cf. 
Appendix~\ref{App}):
\begin{subequations}
  \label{eq:phase-pert-theory} 
\begin{eqnarray}
  \label{eq:MDRC-MDTC}
  \left<\frac{\partial U}{\partial\theta}\right> &=&
    \frac{1}{L}\frac{\varepsilon_m}{\sqrt{\varepsilon_0}}
    \frac{\omega}{2\pi c} \frac{\cos^2\theta}{\cos\theta_0}
    \left| u_0(k) \right|^2 J(q|k),
\end{eqnarray}
with $u_0(k)$ being the polarization dependent Fresnel reflection or
transmission coefficients~\cite{Born} corresponding to the scattering
system of a planar (non-rough) interface, and
\begin{eqnarray}
  \label{eq:integral}
     J(q|k) &=& L e^{-\sigma^2\Lambda^2(q|k)} 
             \int^\infty_{-\infty} du\,
             e^{i(q-k)u} e^{\sigma^2\Lambda^2(q|k) \,W(u)},  
\end{eqnarray}
where
\begin{eqnarray}
  \label{eq:Lambda}
  \Lambda(q|k) &=& \left\{
    \begin{array}{cll}
      \alpha_0(q)+\alpha_0(k), & \qquad & \mbox{in reflection} \\
        \alpha_1(q)-\alpha_0(k), & \qquad & \mbox{in transmission}
    \end{array}
  \right. 
\end{eqnarray}
\end{subequations} 
denotes the momentum transfer perpendicular to the mean surface.  In
writing Eqs.~(\ref{eq:phase-pert-theory}), we recall that $\sigma$ and
$W(u)$ are the surface roughness and correlation function,
respectively (cf.  Eqs.~(\ref{eq:surf-roughness})), and that $L$
denotes the length of the rough surface measured along the mean
surface.  Notice, that for a planar surface ($\sigma= 0$), $J(q|k)$
will be proportional to $\delta(q-k)$ and hence the effect of the
surface roughness in Eq.~(\ref{eq:MDRC-MDTC}) is fully contained in
the $J(q|k)$-integral.

In order to derive an approximate analytic expression to haze, we
start by assuming that $\sigma^2|\Lambda^2(q|k)|\ll 1$ for all values
of $q$ in the radiative region defined by
$\left|q\right| \leq \sqrt{\varepsilon_m}\omega/c$. 
Within this approximation, the $J$-integral takes on the following
form
\begin{eqnarray}
  \label{eq:expansion}
  J(q|k) &\simeq& L e^{-\sigma^2\Lambda^2(q|k)}
       \left[
          2\pi \delta(q-k) + \sigma^2\Lambda^2(q|k) g(|q-k|)
       \right], \quad \sigma^2|\Lambda^2(q|k)|\ll 1,
\end{eqnarray}
where Eq.~(\ref{eq:power-spec}) has been used to relate $W(|x|)$ to
the power spectrum $g(|k|)$. It should be noticed that the first term
of Eq.~(\ref{eq:expansion}) is a coherent (specular) contribution,
while the last term represents the lowest order contribution from the
incoherent (diffuse) field. Moreover, the coherent term depends on the
parameter that defines the surface roughness only through the
rms roughness, $\sigma$, while the incoherent contribution in addition
shows a dependence on the correlation length $a$ {\it via} the power
spectrum $g(|k|)$.

In order to calculate haze, we need to get an expression for the
(integrated) fraction of the incident power that is either reflected
or transmitted into any angle, {\it i.e.} one is looking for an
expression for ${\cal U}$ as defined by Eq.~(\ref{eq:ref}). The main
effect of surface roughness is to alter the angular distribution of
the light being reflected from, or transmitted through, a randomly
rough surface.  However, the amount of integrated reflected or
transmitted light is much less sensitive to the presence of surface
roughness.  Notice, that this is only true if we are not close to a
roughness induced surface resonance like {\it e.g.} those due to the
excitations of surface plasmon polaritons~\cite{SPP}. Another
situation where the above assumption is known to fail is in situations
where the scattered or transmitted intensity is rather low for the
planar surface so that the presence of roughness might renormalize
this result in a significant way. This is for instance the case for
the Brewster angle phenomenon~\cite{Born}.  Hence, for the purpose of
this study, it will be assume that ${\cal U}$ shows little sensitivity
to surface roughness so that it can be well approximated by the planar
result. With Eq.~(\ref{eq:MDRC-MDTC}) and $\sigma=0$ one therefore
has
\begin{eqnarray}
  \label{eq:norm}
  {\cal U} &\simeq& 
        \frac{\alpha_m(k)}{\alpha_0(k)} 
        \left| u_0(k)\right|^2 , 
\end{eqnarray}
where one, as before, should use $m=0$ in reflection and $m=1$ in
transmission.

By substituting Eqs.~(\ref{eq:expansion}) and (\ref{eq:norm}) into
Eqs.~(\ref{eq:phase-pert-theory}) an approximate result for the
$I$-integrals, Eq.~(\ref{eq:I-int}), used to define haze, can be
obtained as
\begin{eqnarray}
  \label{eq:Int-est-tmp}
   I(\theta_-,\theta_+) 
&\simeq& 
\frac{1}{L} \int^{q_+}_{q_-} \frac{dq}{2\pi} \,
     \frac{\alpha_m(q)}{\alpha_m(k)} J(q|k) \nonumber \\
&\simeq& 
e^{-\sigma^2\Lambda^2(k|k)} 
           + \int^{q_+}_{q_-} \frac{dq}{2\pi} \, 
               \sigma^2\Lambda^2(q|k)
               \frac{\alpha_m(q)}{\alpha_m(k)}
                e^{-\sigma^2\Lambda^2(q|k)} g(|q-k|),
\end{eqnarray}
where $q_\pm = k\pm\Delta q$. To try to obtain an estimate for the
last integral of Eq.~(\ref{eq:Int-est-tmp}) will now be approached.
Since the momentum interval $\Delta q$ is small (cf.
Eq.~(\ref{eq:delta-mom})), $\alpha_m(q)$ and $\Lambda(q|k)$ are
therefore slowly varying functions of $q$ in the interval $q_-\leq
q\leq q_+$. 
The power spectrum $g(|q|)$, however, can in
particular for large correlation lengths, $\Delta q a \gg 1$, become
strongly $q$-dependent over the interval of interest even for small
values of $\Delta q$.  Thus, the integrand of the last term of
Eq.(\ref{eq:Int-est-tmp}), accept the power spectrum $g(|p|)$, can be
approximated by its value at $q=k$ (specular direction).  
Hence, one may write
\begin{eqnarray}
  \label{eq:Int-est-tmp-1}
   I(\theta_-,\theta_+) &\simeq& 
     e^{-\sigma^2\Lambda^2(k|k)} 
     \left[ 1+
               \sigma^2\Lambda^2(k|k)
               \frac{G(a) \,\, \Delta q}{\pi}
     \right],
\end{eqnarray}
where 
\begin{eqnarray}
  \label{eq:APS}
  G(a) &=& \frac{1}{2\Delta q}
  \int^{q_+}_{q_-} dq \, g(|q-k|)
     \;=\;
  \frac{1}{2\Delta q}
  \int^{\Delta q}_{-\Delta q}dq \, g(|q|)
\end{eqnarray}
denotes the {\em power spectrum factor} calculated over a momentum
interval of half-width $\Delta q$ around zero momentum transfer
$q-k=0$. It is important to realize that $G(a)$ is independent
of the angle of incidence, but will, of course, depend on the type of
power spectrum used for the surface roughness. Hence, when the type of
power spectrum is known, $G(a)$ can be calculated for {\em all}
angles of incidence, as well as for both reflection and
transmission.  For an exponential power spectrum, defined by
Eq.~(\ref{eq:exp-corr}), the power spectrum factor becomes
\begin{eqnarray}
  \label{eq:APS-exp}
   G(a) &=& \frac{2}{\Delta q}\arctan(\Delta q a),
\end{eqnarray}
and for a Gaussian power spectrum,
$g(|q|)=\sqrt{\pi}a\exp(-q^2a^2/4)$, often used in practice,
$G(a)=(\pi/\Delta q) \, \mbox{erf}(\Delta q a/2)$ where
$\mbox{erf}(\cdot)$ is the error function~\cite{Stegun}. Notice that
whenever $\Delta q a \ll 1$, the power spectrum factor may be expanded
with the result that $G(a)$ is directly proportional to the power
spectrum at zero momentum; $G(a)\simeq g(0)$.  In general,
however, this is not the case.

In the spirit of phase-perturbation theory~\cite{Shen,SG,Rosa}, one
observes that the terms in square brackets in
Eq.~(\ref{eq:Int-est-tmp-1}) are the first few (non-trivial) terms of
an exponential function.  Hence, one may write
\begin{eqnarray}
  \label{eq:Int-est-tmp-2}
   I(\theta_-,\theta_+) &\simeq& 
     \exp\left[-\sigma^2\Lambda^2(k|k)
              \left(1-\frac{G(a) \Delta q}{\pi} \right)
          \right], 
\end{eqnarray}
and after substituting this expression back into
Eq.~(\ref{eq:haze-B}), the following approximate expression for haze
is finally arrived at
\begin{eqnarray}
  \label{eq:haze-final}
  {\cal H}(\theta_0) &=& 
        1-\exp\left[-\sigma^2\Lambda^2(k|k)
              \left(1-\frac{G(a) \Delta q}{\pi} \right)
          \right] . 
\end{eqnarray}
Here the perpendicular momentum transfer, $\Lambda(q|k)$, has already
been defined in Eq.~(\ref{eq:Lambda}), the power spectrum
factor $G(a)$  in Eq.~(\ref{eq:APS}), and the momentum
interval $\Delta q$ in Eq.~(\ref{eq:delta-mom}).  The
approximation~(\ref{eq:haze-final}) and its numerical confirmation (to
be presented later), are the main results of this study.

There are several important observations to be made from the
approximate expression to haze~( \ref{eq:haze-final}):
First, it should be observed that according to
Eq.~(\ref{eq:haze-final}) haze can be written in terms of {\em two
dimensionless quantities}:
$\sigma\Lambda(k|k)$ and $G(a) \Delta q$.  The product of the
rms-roughness of the surface topography and the perpendicular momentum
transfer of the reflection (or transmission) process,
$\sigma\Lambda(k|k)$, does depend on the ``amount'' of roughness {\em
  but not} on how it is being correlated.  The quantity, $G(a)
\Delta q$, on the other hand, depends on the power spectrum, and is
therefore sensitive to the type of height-height correlation function
respected by the rough surface.  Second, the dependence on the angle
of incidence {\em only} enters through the momentum transfer:
$\sigma\Lambda(k|k)$ (perpendicular) and $\Delta q$ (lateral).  Third,
to apply the approximation~(\ref{eq:haze-final}) in practical
applications, it is the power spectrum of the surface roughness around
zero momentum that is of interest since only this portion of the power
spectrum enters the definition of the power spectrum factor $G(a)$. This is important to realize since the whole power spectrum,
and in particular its tail, is often difficult to assess in a reliable
way from direct measurements of the surface topography~\cite{Ogilvy}.

Another quantity used frequently by the optical industry to quantify
visual appearance of surfaces are, as mentioned earlier,
gloss~\cite{Gloss}. Gloss is, crudely speaking, a measure of how
specular a surface appears in reflection or transmission. This can be
measured by {\it e.g.} the fraction of the reflected (transmitted)
light that is reflected (transmitted) into in a small angular interval
around the specular direction $\Theta$. Hence, gloss is more or less
complementary to haze, {\it i.e.} it can more or less be written as
$1-{\cal H}(\theta_0)$. Hence, the last term of
Eq.~(\ref{eq:haze-final}) may therefore be considered as an expression
for gloss. In practice one talks of at least two types of gloss; wide
angle gloss and specular gloss~\cite{Gloss}.  They are distinguished
by the values of $\Delta \theta$ (and therefore $\Delta q$) used to
define them. When using the expression for gloss described above, one
has to substitute the correct values for $\Delta \theta$ (as well as
$\theta_0$). The expression for gloss, $1-{\cal H}(\theta_0)$, is
probably best suited for specular gloss due to the approximation
introduced in order to arrive at Eq.~(\ref{eq:Int-est-tmp-1}). However
for large values of $\Delta \theta$ it is not expected to work too
well.

Previously, Alexander-Katz and Barrera~\cite{Barrera}, while studying
gloss (in reflection), found that the reduced variables for this
problem were $(\sigma/\lambda)\cos\theta_0$ and
$(a/\lambda)\cos\theta_0$. The results of these authors, have quite a
few common features with those presented in this study. However, this
is probably not so surprising since gloss and haze can be viewed as
more or less complementary quantities. In particular, in reflection,
we find that $\sigma\Lambda(k|k)$ indeed scales as
$(\sigma/\lambda)\cos\theta_0$ (cf. Eq.(\ref{eq:haze-ref}) below).
Hence, this agrees with the result found by Alexander-Katz and Barrera
for gloss~\cite{Barrera}. However, for the correlation function
dependent quantity, we do not in general seem to agree with these
authors.  Only in the limit $\Delta q a\ll 1$, when $G(a)\simeq g(0)a$
as mentioned earlier, do we tend to get the same scaling behavior as
reported previously by Alexander-Katz and Barrera. The reason for this
``discrepancy'', is related to the level of accuracy applied in the
approximation of the integral of Eq.~(\ref{eq:Int-est-tmp}).

When $|\sigma\Lambda(k|k)|\ll 1$, the exponential function of
Eq.~(\ref{eq:haze-final}) can be expanded with the result that
\begin{eqnarray}
  \label{eq:haze-expansion}
  {\cal H}(\theta_0) &\simeq& \sigma^2\Lambda^2(k|k)
              \left(1-\frac{G(a) \Delta q}{\pi} \right) , \qquad
  |\sigma\Lambda(k|k)|\ll 1.
\end{eqnarray}
This has the consequence that the ratio of haze in reflection~(${\cal
  H}_s(\theta_0)$) and transmission~(${\cal H}_t(\theta_0)$) will be
independent of the rms-roughness $\sigma$ and only depend on the
correlation length $a$ (and the form of the power spectrum) as well as
the parameters defining the scattering geometry.

For completeness, the expressions for haze in terms of the
``defining'' quantities will be explicitly given. With
Eqs.~(\ref{eq:delta-mom}), (\ref{eq:Lambda}) and (\ref{eq:APS}) one
has for reflection
\begin{subequations}
  \begin{eqnarray}
    \label{eq:haze-ref}
     {\cal H}_s(\theta_0) &\simeq&  
        1-\exp\left[
            -16\pi^2 \varepsilon_0 \left(\frac{\sigma}{\lambda}\right)^2\cos^2\theta_0\,  
            \left\{1- 2\sqrt{\varepsilon_0}
                   \frac{G(a)}{\lambda}
                   \sin\Delta\theta   
                  \cos\theta_0 
                 \right\}
            \right],
  \end{eqnarray}
  while in transmission the haze can be approximated by
  \begin{eqnarray}
    \label{eq:haze-trans}
     {\cal H}_t(\theta_0) &\simeq&  
       1 - \exp\left[
     -4\pi^2 \varepsilon_0 \left( \frac{\sigma}{\lambda}\right)^2
  \left\{\sqrt{\frac{\varepsilon_1}{\varepsilon_0} -\sin^2\theta_0}
        -\cos\theta_0 \right\}^2
  \left\{1 - 2\sqrt{\varepsilon_1}
                   \frac{G(a)}{\lambda}
                   \sin\Delta\theta   
                   \sqrt{1-\frac{\varepsilon_0}{\varepsilon_1}\sin^2\theta_0}
             \right\}
        \right]. 
  \end{eqnarray}
\end{subequations}
In obtaining Eq.~(\ref{eq:haze-trans}) it has been used that for the
specular direction (in transmission)
$\cos\Theta_t=\sqrt{1-(\varepsilon_0/\varepsilon_1)\sin^2\theta_0}$.

Before closing this section, it should be stressed that the
approximate expression to haze, Eq.~(\ref{eq:haze-final}) and
related expressions, are based on phase-perturbation theory. Hence,
its validity will become questionable when multiple scattering starts
to contribute significantly to the scattered or transmitted fields.
We stress that phase-perturbation theory is not a small amplitude
perturbation theory, so it may still give reliable results for
strongly rough surfaces in the large correlation length limit ({\it
  i.e.} small slopes).  The accuracy of Eq.~(\ref{eq:haze-final}) will
be investigated in Sec.~\ref{Sec:Results}.

\subsection{Gloss} 

In addition to haze, {\it gloss} is a quantity frequently used in the
optical industry to quantify optical materials.  As haze quantifies
the fraction of the reflected or transmitted light that is directed
outside a given angular interval, gloss, on the other hand, is related
to the amount of light that falls inside the same interval. 

Mathematically, gloss, ${\cal G}(\theta_0)$, is defined as the integral
$I(\theta_+,\theta_-)$ of Eq.~(\ref{eq:I-int}) with $\theta_\pm$
adapted to the particular definition of gloss of interest.  Hence, in
terms of the surface parameters, gloss can be expressed by
Eq.~(\ref{eq:Int-est-tmp}), or by the subsequent approximate
expressions that followed it. In particular take notice of
Eq.~(\ref{eq:Int-est-tmp-2}) that represents an approximate formula
for gloss
\begin{eqnarray}
 \label{eq:gloss-final}
 {\cal G}(\theta_0)
    &\simeq& 
    \exp\left[-\sigma^2\Lambda^2(k|k)
              \left(1-\frac{G(a) \Delta q}{\pi} \right)
          \right]. 
\end{eqnarray} 
In light of Eq.~(\ref{eq:haze-final}), one, as expected, observes that
${\cal H}_i(\theta_0)+{\cal G}_i(\theta_0)=1$, where
$i=s(\mbox{reflection}), t(\mbox{transmission})$ and $\theta_\pm$ are
the same for both haze and gloss. This means that when the value of
haze is increasing, gloss is reduced and visa versa.  We stress that
both quantities in the equation above refer to reflection or
transmission. Mathematically there is no problem in consider gloss in
transmission, however, this term is not commonly used in practical
applications.\footnote{Gloss is normally only used in the context of
  reflection. Furthermore, the angles of incidence that haze and gloss
  refer to, are often different (and standard dependent).} For the
implications and validity of the above approximate expression to
gloss, Eq.~(\ref{eq:gloss-final}), the interested reader is referred
to a separate publication~\cite{our-gloss-paper}.

\section {Simulation Results and Discussion}
\label{Sec:Results}

\subsection {Comparison to Monte Carlo simulations}

The approximate expression for haze, Eq.~(\ref{eq:haze-final}), is
based on phase-perturbation theory, and is therefore not rigorous.  In
order to investigate how well this analytic approximation is
performing, and when this scaling form breaks down, it will be
compared to what can be obtained from a rigorous computer simulation
approach~\cite{AnnPhys,ThesisIngve}.  Such an approach is formally
exact, since it solves the Maxwell equations numerically without
applying any approximations. It will thus take into account any higher
order scattering process, and not just those accounted for by
phase-perturbation theory. 

Such exact Monte Carlo simulations can be performed by formulating the
Maxwell equations as a coupled set of integral equations. This is done
by taking advantage of Green's second integral identity in the plane
as well as the boundary conditions satisfied by the fields and their
normal derivative on the randomly rough surface.  These integral
equations can be converted into matrix equations and solved for the
sources --- the fields and their normal derivative evaluated at the
surface. From the knowledge of these sources, the scattered
(transmitted) field at any point above (below) the surface may be
calculated and therefrom the mean DRC (DTC). The whole detailed
procedure for doing such Monte Carlo simulations can be found in
Refs.~\cite{AnnPhys,ThesisIngve,Wavelength_dependence}.

\begin{figure}
  \centering  
  \includegraphics*[width=8cm]{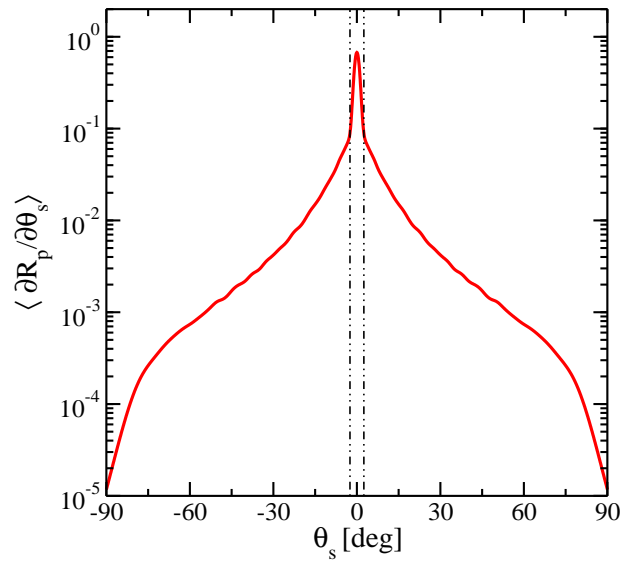}
  \caption{(Color online) The mean differential reflection
    coefficient, $\left<\partial R_p/\partial \theta_s\right>$, {\it
      vs.} the scattering angle $\theta_s$ as obtained by rigorous
    Monte Carlo simulations. The incident $p$-polarized light of
    wavelength $\lambda=0.6328 \mu m$ was incident at normal
    incident~($\theta_0=0^\circ$) onto the rough dielectric surface,
    and for the incident wave a finite sized beam of half-width
    $g=6.4\, \mu m$ was used in order to reduce end effects.  This
    surface of length $L=25.6 \mu m=40.5\lambda$ separates vacuum,
    above the surface, from the dielectric medium of dielectric
    constant $\varepsilon_1(\omega)=2.25$ below the surface.The
    statistical properties of the randomly rough surface was
    characterized by a Gaussian height distribution function of (rms)
    width $\sigma=0.037 \mu m= 0.058\lambda$ and an exponential
    correlation function of correlation length $a=1 \mu m=
    1.58\lambda$.  The surface was discretized at $N_\zeta=500$
    equally distributed points, and the result was averaged over
    $N_\zeta=5000$ surface realizations. The vertical dash-dotted
    lines are at $\theta_\pm=\pm2.5^\circ$. Notice the specular
    (coherent) peak around $\theta_s=\theta_0$. The haze in reflection
    for this surface is according to the numerical simulations ${\cal
      H}(\theta_0)=0.33$, while the prediction of
    Eq.~(\protect\ref{eq:haze-final}) is ${\cal H}(\theta_0)=0.34$. }
    \label{fig:MDRC-ex} 
\end{figure}

In the numerical simulation results for haze to be presented below,
one will first calculate $\left<\partial U/\partial \theta\right>$ and
then, by numerical integration, calculate haze directly from
Eq.~(\ref{eq:haze}). An example of a mean differential reflection
coefficient curve, obtained by numerical simulations for a polymer
material, is given in Fig.~\ref{fig:MDRC-ex}. In obtaining this result
one assumed a Gaussian height distribution function of
$\sigma/\lambda=0.058$ and an exponential correlation function of
correlation length $a=1.58\lambda$. These parameters are similar to
those found for the measured surface of
Fig.~\ref{fig:experimental-surf-data}.  The vertical dash-dotted lines
that can be seen in Fig.~\ref{fig:MDRC-ex} are at an angular position
$\theta_\pm=2.5^\circ$ --- {\it i.e.} at the angles about the specular
direction $\theta_s=\theta_0=0^\circ$ used in the definition of haze.
By numerical integration one may from this result calculate haze, and
for this particular example one finds ${\cal
  H}_s(\theta_0=0^\circ)=0.34$.

\begin{figure}
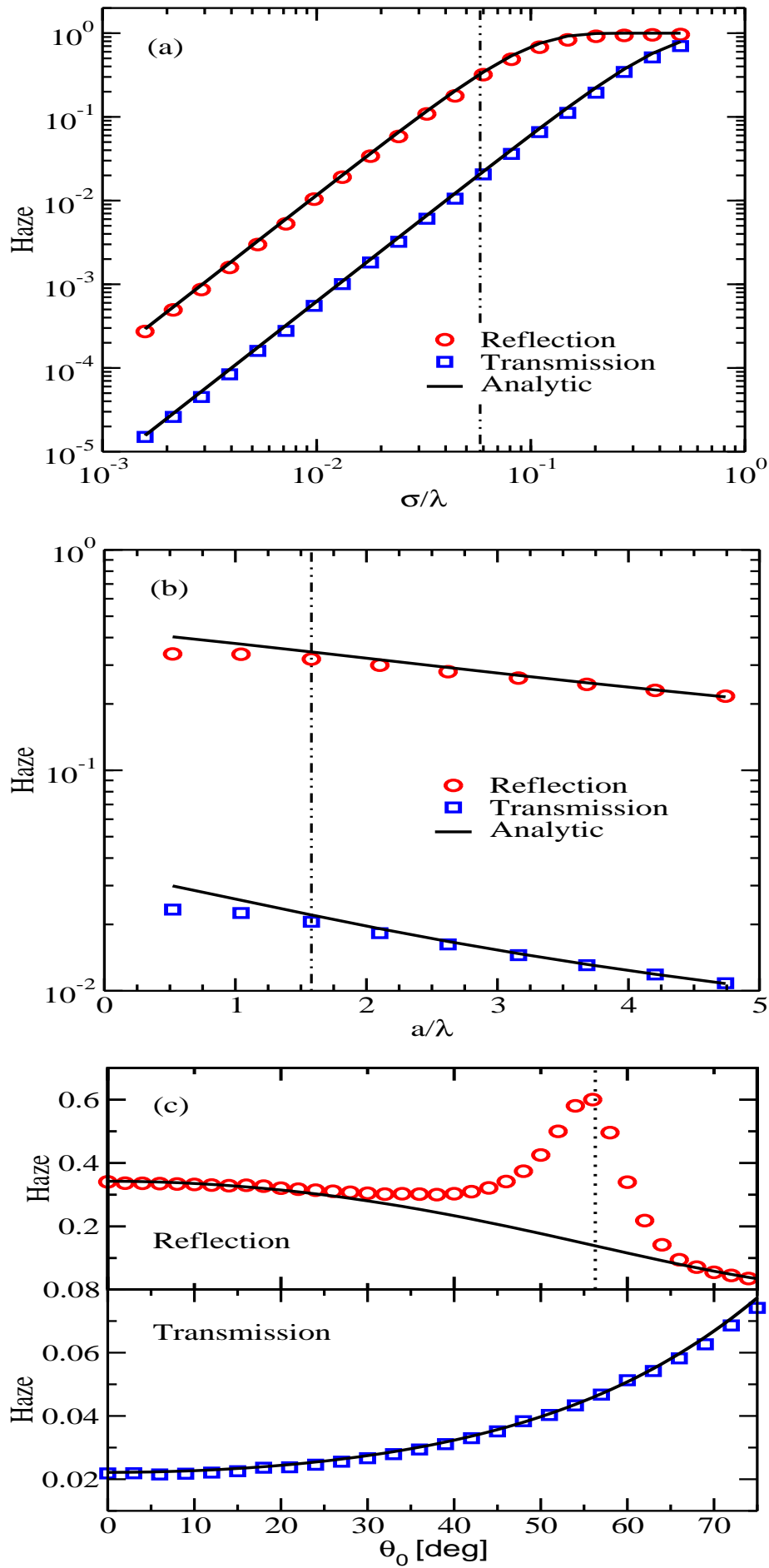

  \centering
  \includegraphics*[width=12cm,height=8cm]{haze_vs_rms-100.eps} \\*[0.4cm]
  \includegraphics*[width=12cm,height=8cm]{haze_vs_corr-037.eps}\\*[0.4cm]
  \includegraphics*[width=12cm,height=8cm]{haze_vs_angle-100-0.37.eps}
\caption{(Color online) Haze, ${\cal H}$, as a function of (a) the surface
  roughness $\sigma/\lambda$ for $a/\lambda=1.58$ at normal
  incidence, (b) the correlation length $a/\lambda$ for
  $\sigma/\lambda=0.058$  at normal incidence, and (c) the angle
  of incidence $\theta_0$ for $\sigma/\lambda=0.058$ and
  $a/\lambda=1.58$. For all figures the wavelength of the $p$-polarized
  incident light was $\lambda=0.6328 \, \mu m$.  The open symbols are
  results of rigorous Monte Carlo simulations, while the solid lines
  are the predictions of Eq.(\protect\ref{eq:haze-final}).  The
  dashed-dotted line in the upper panel (reflection) of
  Fig.~\protect\ref{fig:Haze-scaling}(c) corresponds to the position of
  the Brewster angle, $\theta_0=\theta_B$ determined by
  $\tan^2\theta_B=\varepsilon_1/\varepsilon_0$. 
  }
    \label{fig:Haze-scaling} 
\end{figure}

In Figs.~\ref{fig:Haze-scaling} rigorous numerical simulation results
for haze (open symbols) are presented versus various parameters of the
scattering geometry in both reflection and transmission.  Here an
exponential correlation function $W(|x|)=\exp(-|x|/a)$, that is a
reasonable fit to measured data for some polymer films~(see
Figs.~\ref{fig:experimental-surf-data}) has been assumed.
Furthermore, the light of wavelength $\lambda=0.6328\mu m$ was
incident normally onto the mean
surface~(Figs.~\ref{fig:Haze-scaling}(a) and (b)), and for the
dielectric constant of the media involved, $\varepsilon_0=1$ and
$\varepsilon_1=2.25$, were used.  For all simulations, the results
were averaged over at least $N_\zeta=500$ surface realizations.  The
vertical dashed-dotted lines present in
Figs.~\ref{fig:Haze-scaling}(a) and (b) correspond to the roughness
parameters used in obtaining the results of Fig.~\ref{fig:MDRC-ex}.
The solid lines in Figs.~\ref{fig:Haze-scaling} are the predictions of
the Eq.~(\ref{eq:haze-final}) --- the approximate expression to
haze.  In particular, Fig.~\ref{fig:Haze-scaling}(a) presents the
dependence of haze {\it vs.}  $\sigma/\lambda$ for a fixed value of
the correlation length $a/\lambda=1.58$ and the angle incidence was
$\theta_0=0^o$. This value for $a/\lambda$ corresponds to the vertical
dashed-dotted line in Fig.~\ref{fig:Haze-scaling}(b).  From
Fig.~\ref{fig:Haze-scaling}(a) it is observed that the analytic
approximations (solid lines) performs impressing well, also for the
roughest surfaces considered.  It should be noted that both the
$(\sigma/\lambda)^2$ increase of haze for low levels of roughness (cf.
Eq.~(\ref{eq:haze-expansion})) as well as the ``bend-off'', or
saturation, that takes place for the haze for strong roughness seem to
be correctly predicted by Eq.~(\ref{eq:haze-final}). In this latter
case, however, the numerical simulation results saturated around
${\cal H}\simeq0.95$--$0.96$ while the theoretical curves approach the
value of one. Hence, for strongly rough surfaces one has a relative
error of about $5\%$ as discussed in an earlier section.

In Fig.~\ref{fig:Haze-scaling}(b) the dependence of haze {\it vs.} 
correlation length $a/\lambda$ for $\sigma/\lambda=0.058$ (vertical
dashed-dotted line) and $\theta_0=0^o$ is depicted.  The agreement,
also in this cases, is rather satisfactory. However, from this figure
there is an indication that at smaller correlation lengths, the
agreement becomes less good. This is caused by phase-perturbation
theory not being a good approximation in the small correlation length
limit (that corresponds to large local slopes $\sigma/a$).  

Finally in Figs.~\ref{fig:Haze-scaling}(c), how the angle of incidence
influence haze is studied. Only positive angles of incidence are being
considered since the scattering geometry is so that there is a
symmetry with respect to a change of sign in $\theta_0$. As was done
to obtain the results of Figs.~\ref{fig:Haze-scaling}(a) and (b), one
has also here fixed the roughness and correlation length to the values
used in obtaining the results of Fig.~\ref{fig:MDRC-ex}, {\it i.e.},
$\sigma/\lambda=0.058$ and $a/\lambda=1.58$. In the case of
transmission (lower panel of Fig.~\ref{fig:Haze-scaling}(c)), the
agreement between the Monte Carlo simulation result and the analytic
approximation is of a good quality for all angle of incidence
considered. In reflection (upper panel of
Fig.~\ref{fig:Haze-scaling}(c)), the agreement is of a good quality
only for the smallest scattering angles. However, as the angle of
incidence approaches roughly $\theta_0=55^o$, from below or above, the
disagreement between the simulation and approximate result (for
reflection) becomes pronounced. The reason for this discrepancy is the
so-called Brewster angle phenomenon~\cite{Born}.  This phenomenon
express itself for the planar geometry in $p$-polarization by the
reflection coefficient being exactly zero at the Brewster angle
$\theta_B$ defined by $\tan^2\theta_B=\varepsilon_1/\varepsilon_0$.
However, as roughness is introduced into the system, the reflectivity
of the (rough) surface will not be zero any more, not even at the
Brewster angle, but will instead go through a minimum for an angle of
incidence close to $\theta_B$ (the ``quasi''-Brewster angle
phenomenon). This has the consequence that haze (in reflection and for
p-polarization) will go through a corresponding maximum for the same
angle of incidence.  So in the region about the Brewster angle, the
presence of surface roughness will strongly renormalize the
corresponding planar geometry result with the consequence that the
integrated reflected energy is not any more well approximated by
Eq.~(\ref{eq:norm}). By taking into account roughness in the
estimation of the total integrated scattered intensity, one will most
likely be able to also predict the behavior of haze for such angles of
incidence with more confidence. However, the penalty for doing so, is
that the resulting expressions become much more complicated and must
be evaluated numerically.  Since this is not the aim of the present
study, we will not follow up this line of actions here, but only keep
in mind that the simple approximation (\ref{eq:haze-final}) breaks
down around the Brewster angle. Notice that there is no Brewster angle
phenomenon in transmission as shown explicitly in the lower panel of
Fig.~\ref{fig:Haze-scaling}(c), nor is there any such phenomenon in
reflection (or transmission) for $s$-polarized incident light.  Hence,
for $s$-polarization, the approximate expression for haze,
Eq.~(\ref{eq:haze-final}), should apply for all angles of incidence,
something that has been confirmed by numerical simulations (results
not shown).

\begin{figure}
  \subfigcapskip -3mm
  \centering
  \subfigure[Rigorous simulation result in reflection]{
    \includegraphics*[width=0.45\textwidth,height=0.45\textwidth,angle=-90]{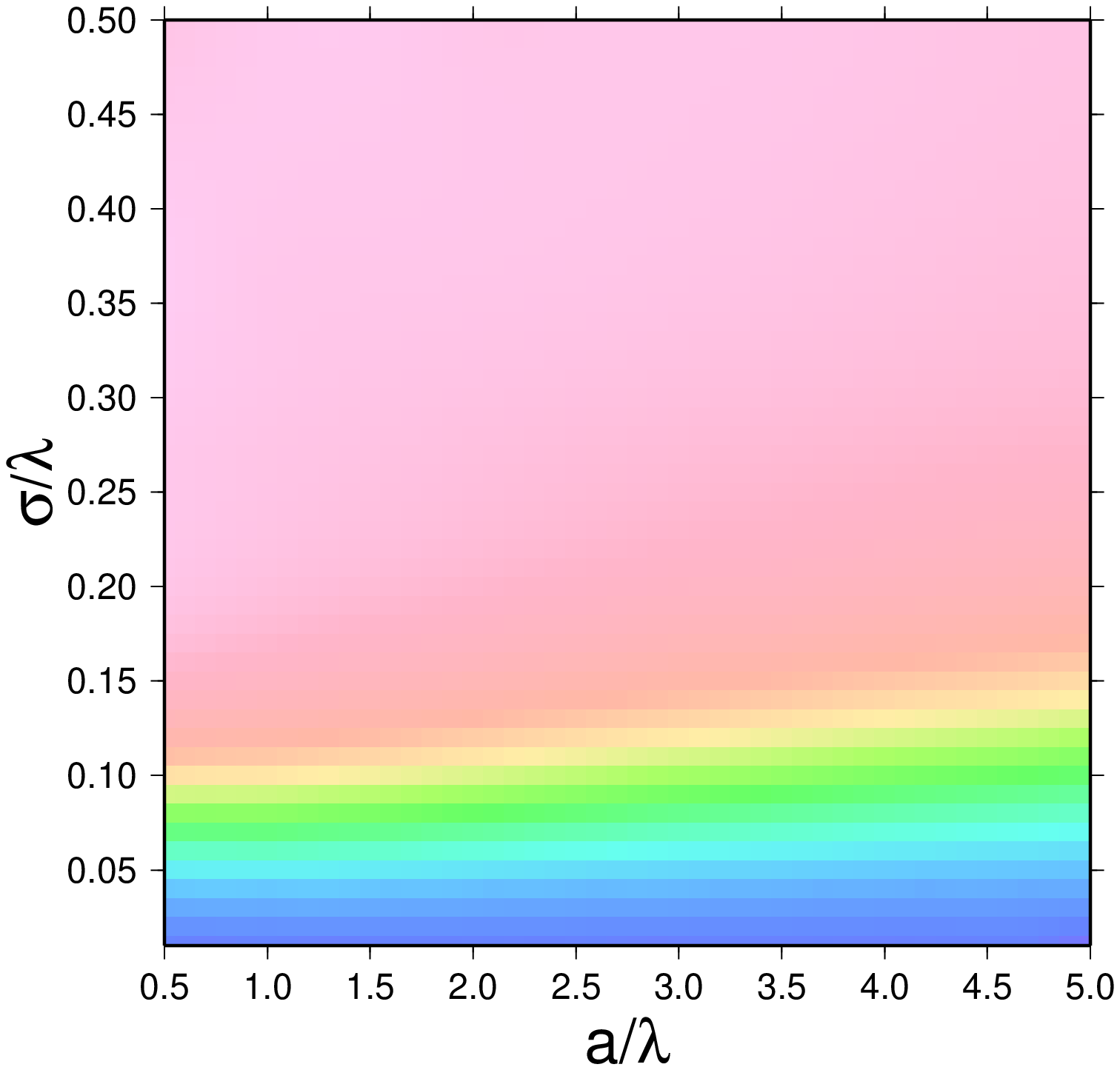} }\quad \quad
  \subfigure[Analytic approximation in reflection]{
    \includegraphics*[width=0.45\textwidth,height=0.45\textwidth,angle=-90]{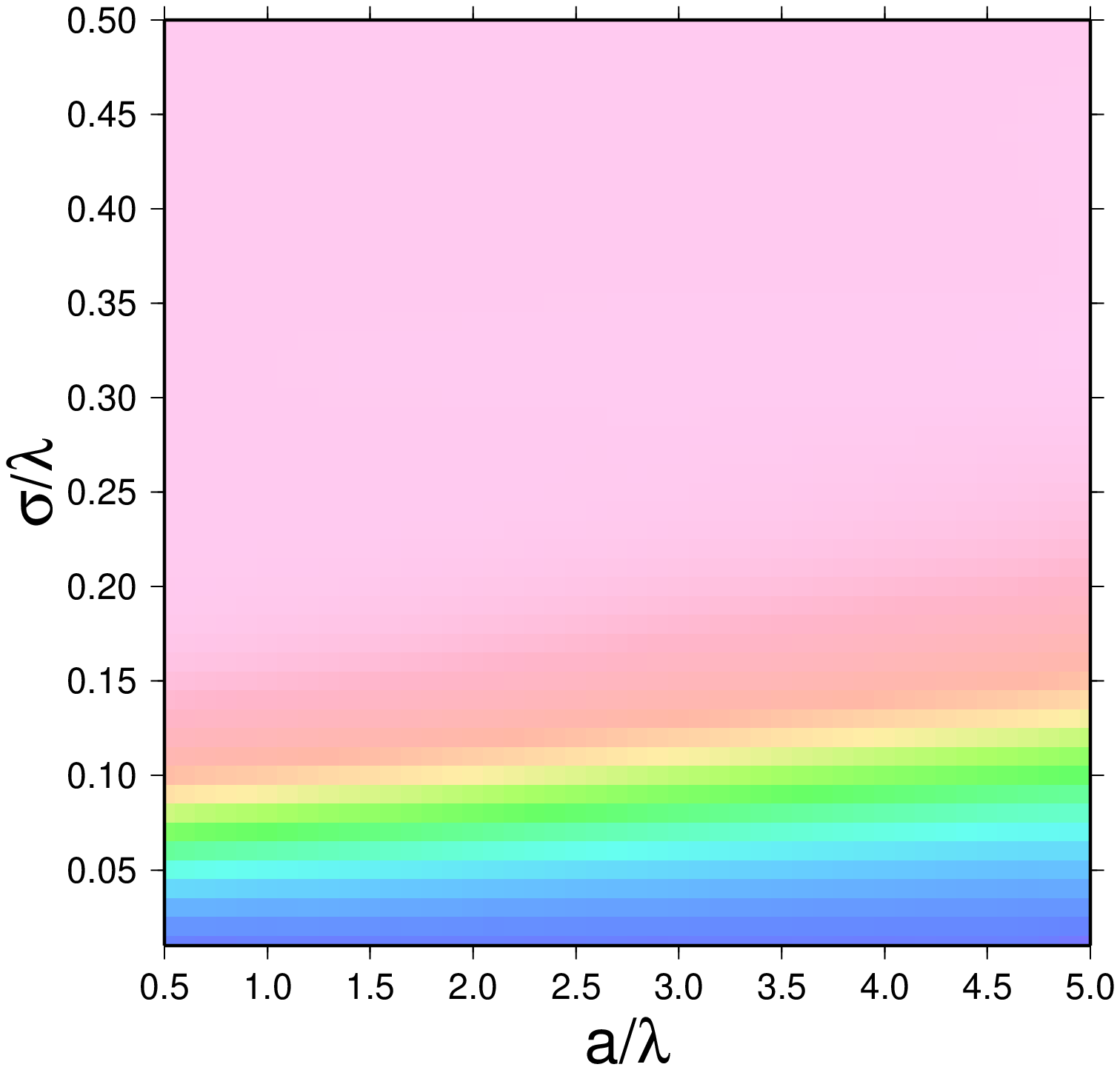} } 
\\*[3mm]
  \subfigure[Rigorous simulation result in transmission ]{
    \includegraphics*[width=0.45\textwidth,height=0.45\textwidth,angle=-90]{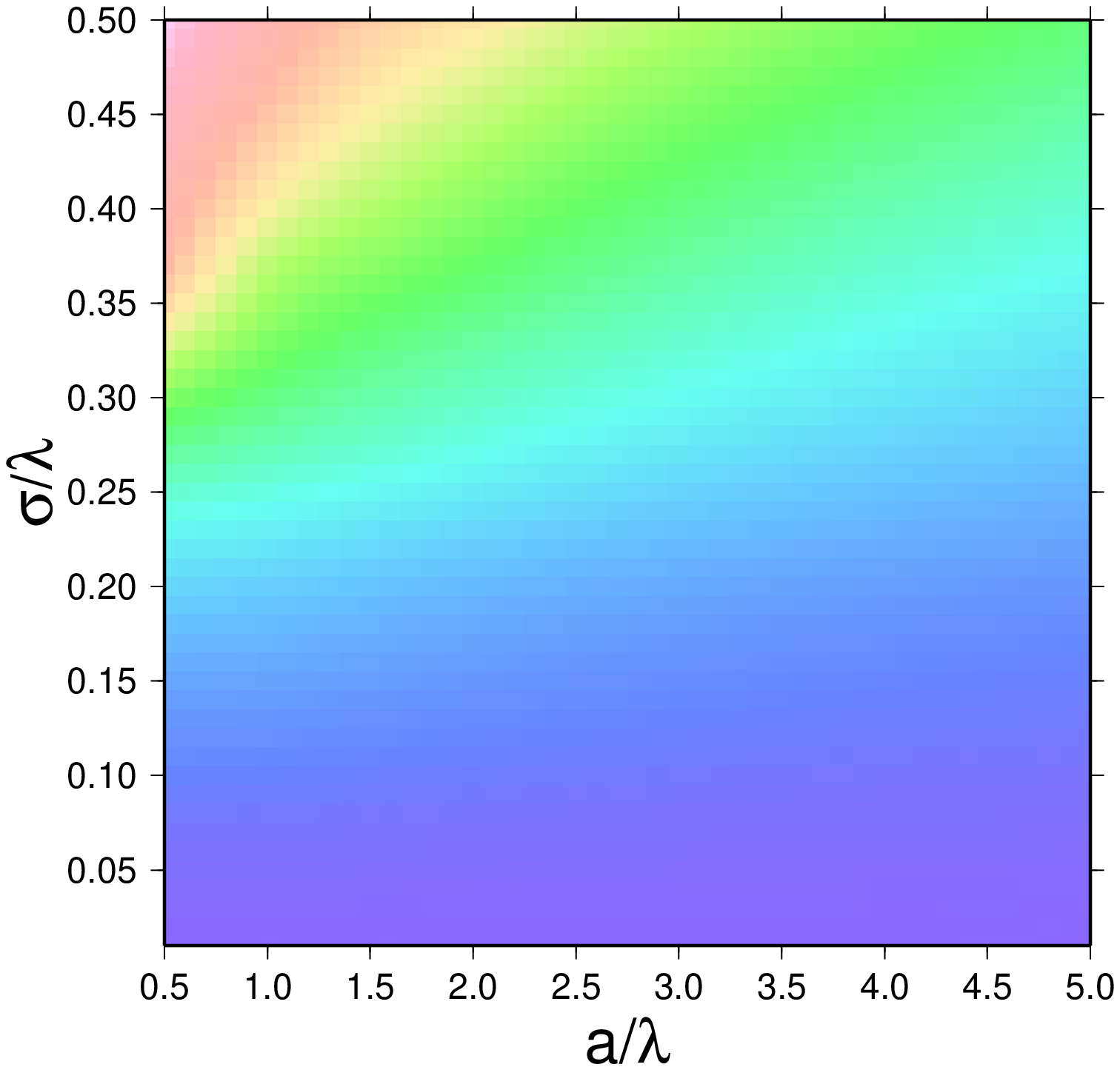}}\quad \quad
  \subfigure[Analytic approximation in transmission]{
  \includegraphics*[width=0.45\textwidth,height=0.45\textwidth,angle=-90]{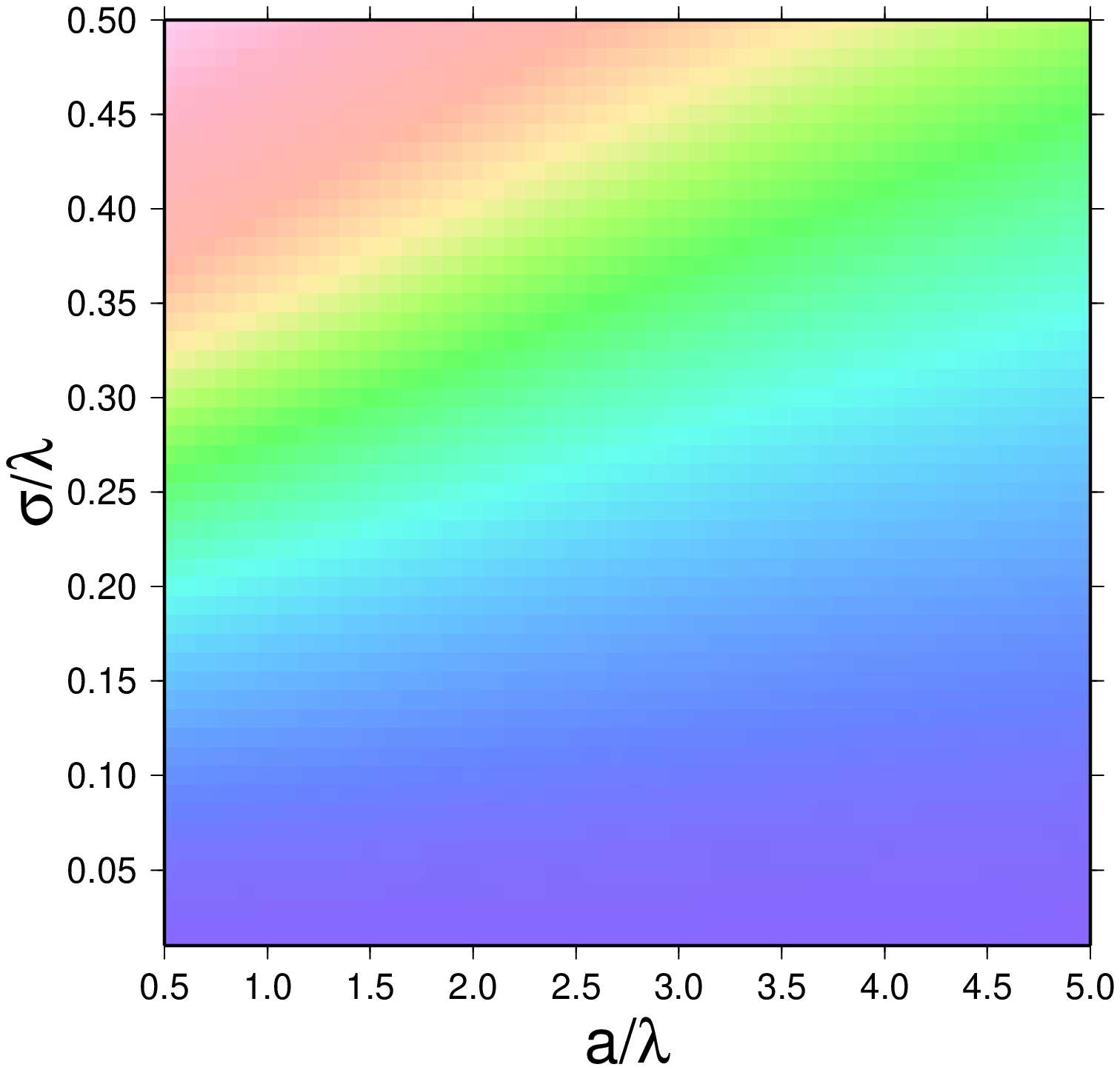} } 
  \\*[5mm]
  \includegraphics*[height=0.5\textwidth,angle=-90]{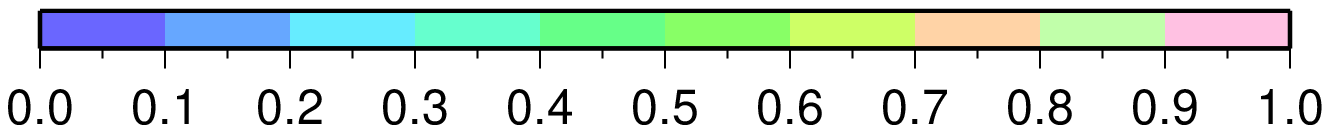} 
\caption{(Color online) Contour plots of the rigorous Monte Carlo simulation
  results~(Figs.~\protect\ref{fig:Contour}(a) and (c)) and the
  analytic approximation
  (\protect\ref{eq:haze-final})~(Figs.~\protect\ref{fig:Contour}(b)
  and (d)) for haze obtained in reflection and transmission for light
  of wavelength $\lambda=0.6328\mu$m incident
  normally~($\theta_0=0^\circ$) onto the rough surface.  The
  dielectric media that was separated from vacuum by a rough
  interface, was characterized by the dielectric constant
  $\varepsilon_1=2.25$.  All results were averaged over {\em at least}
  $N_\zeta=500$ surface realizations. The random surfaces were all
  characterized by a Gaussian height distribution function of standard
  deviation $\sigma$, and an exponential height-height correlation
  function of correlation length $a$. Overall the agreement between
  the analytic and rigorous simulation results is satisfactory over
  large regions of parameter space.}
   \label{fig:Contour}
\end{figure}

In Figs.~\ref{fig:Haze-scaling}(a) and (b), one, or more, of the
parameters that characterize the surface roughness were fixed to
constant values. It is, however, important to get a more complete
picture of the quality of our analytic expression to haze. This can be
achieved by allowing both $\sigma/\lambda$ and $a/\lambda$ to vary
freely (within certain limits).  Figs.~\ref{fig:Contour} depict
contour plots for the variations in haze {\it vs.} both of the two
above mentioned parameters in reflection and transmission. The figures
in the left column, Figs.~\ref{fig:Contour}(a) and (c), show contour
plots of haze as obtained by rigorous Monte Carlo simulations.  In the
right column, {\it i.e.} in Figs.~\ref{fig:Contour}(b) and (d), the
corresponding plots obtained from the analytic haze (approximate)
expression, Eq.~(\ref{eq:haze-final}), are presented.  By in
Figs.~\ref{fig:Contour} comparing the numerical simulation results to
those of the corresponding analytic predictions, one can conclude that
the quality of the analytic expression~(\ref{eq:haze-final}) is
remarkably good over large regions of parameter space. This is
particularly the situation when considering reflection. 
As a general trend, it
seems fair to say that the approximation~(\ref{eq:haze-final})
performs the best for large correlation lengths and smallest
rms-roughness. This is in particular the case for haze in transmission
for which the approximation is poorer than in reflection.
These findings fit the picture that phase-perturbation theory can be
looked upon as a generalization of the more familiar Kirchhoff
approximation that is known to work the best in the large $a/\lambda$
limit~\cite{Beckmann,Ogilvy}. Moreover, notice that when the large
correlation length limit is taken, for fixed rms-roughness, the small
slope limit is approached since the average slope of the surface is
proportional to $\sigma/a$.

So far, a semi-infinite dielectric transparent medium bounded to
vacuum by a randomly rough interface has been considered. From a
practical point of view, it will be rather interesting to also
investigate how haze depends on the surface parameters for, say, a
film geometry. For such a case, the derivation of an approximate
expression for haze, analogous to those presented in
Sec.~\ref{Sec:Haze} for a semi-infinite medium, is lengthy, but can be
performed.  However, such expressions will not be presented here. 
Instead we would like to add that rigorous numerical simulations for a
film geometry have been preformed for normal incidence. The results of
such simulations show that the characteristic dependence of haze on
the parameters $\sigma/\lambda$ and $a/\lambda$, originally found for
a semi-infinite randomly rough surface also seems to hold true for the
film geometry.

\subsection {Comparison to experimental results}

So far in this paper we have mainly considered one-dimensional
surfaces.  However, naturally occurring surfaces, as well as man-made
surfaces generated by, say, an industrial process, are usually
two-dimensional.  Neither is it not uncommon that their statistical
properties are not well described by {\em simple} mathematical
distribution and correlation functions of the form often assumed in
theoretical studies. It is therefore an open question if the
approximate result for haze, Eq.~(\ref{eq:haze-final}), obtained for
one {\em single} randomly rough one-dimensional surface has any
relevance for the more complicated scattering systems encounter in
practical applications. In the most general case our approximate
expression is obviously not suitable. One may still hope, however,
that the general behavior of haze found in Eq.~(\ref{eq:haze-final})
could be taken over to higher-dimensional and more complicated
scattering geometries. In particular, for weakly rough isotropic
surfaces at normal incidence there are hopes that a one-dimensional
approach might work reasonably well. This is so since under such
circumstances the (two-dimensional) mean differential reflection and
transmission coefficients are rotational symmetric, {\it i.e.}  no
$\phi$-dependence. Consequently, a one-dimensional mean differential
or transmission coefficient might be enough to catch the main angular
dependence of the scattering up to cross-polarization effects.  The
purpose of this sub-section is to look into these questions, and to
compare experimental measurements with what can be obtained from a one
dimensional computer simulation approach of the type applied
previously in this paper.

\begin{figure}
  \centering
      \includegraphics*[width=8cm]{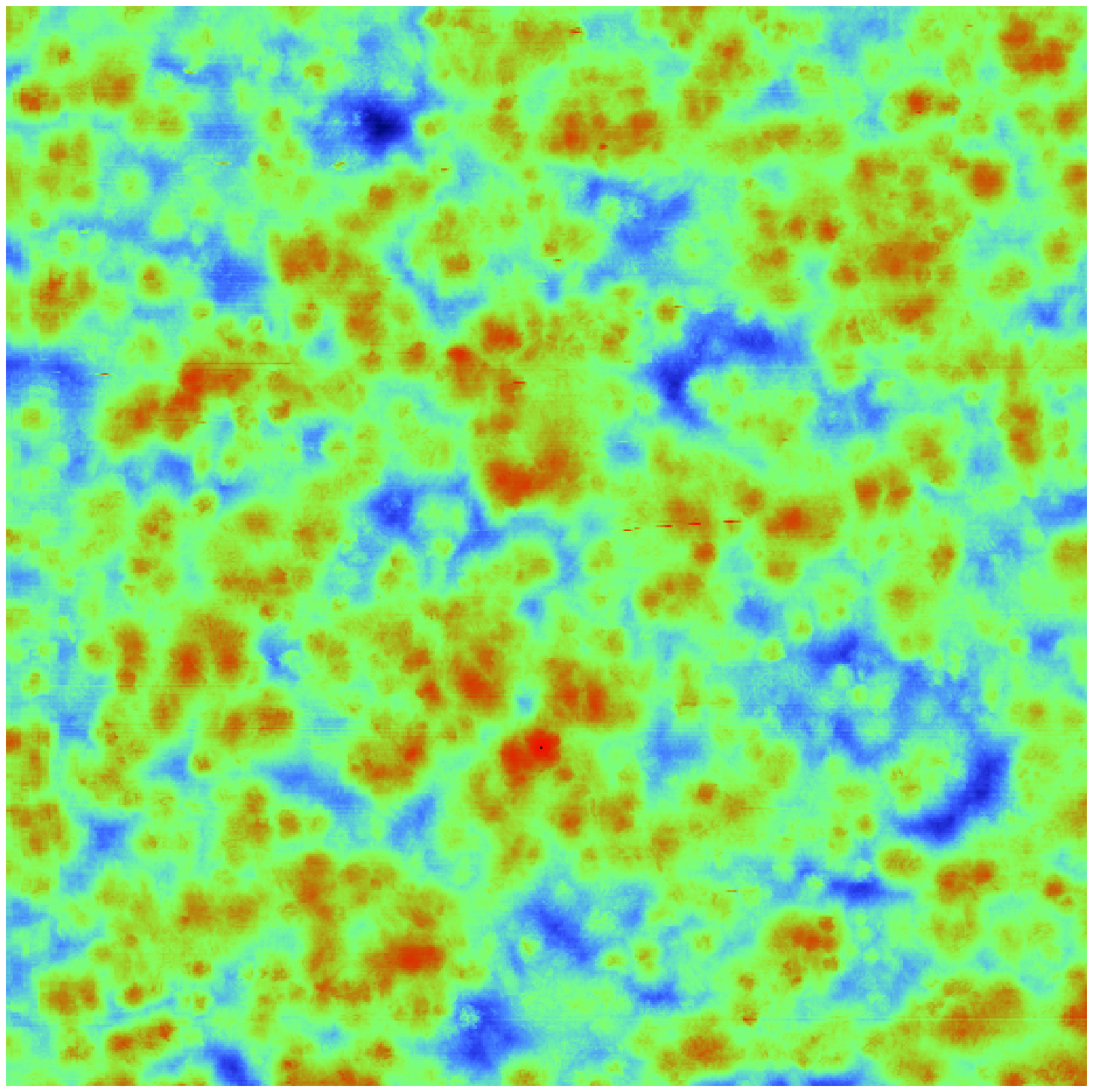}\\*[0.5cm]
      \includegraphics*[width=9cm]{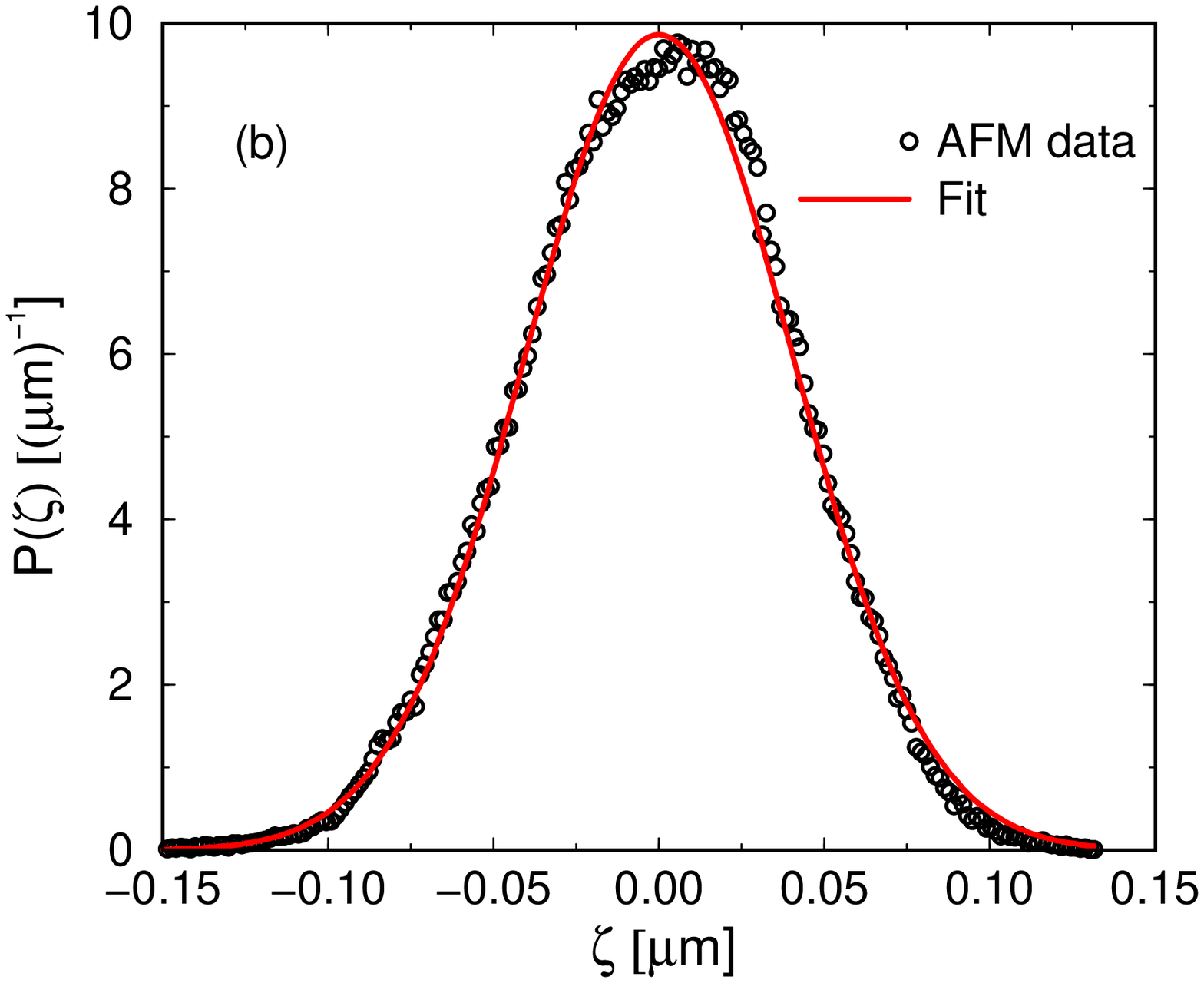}   \\
      \includegraphics*[width=9cm]{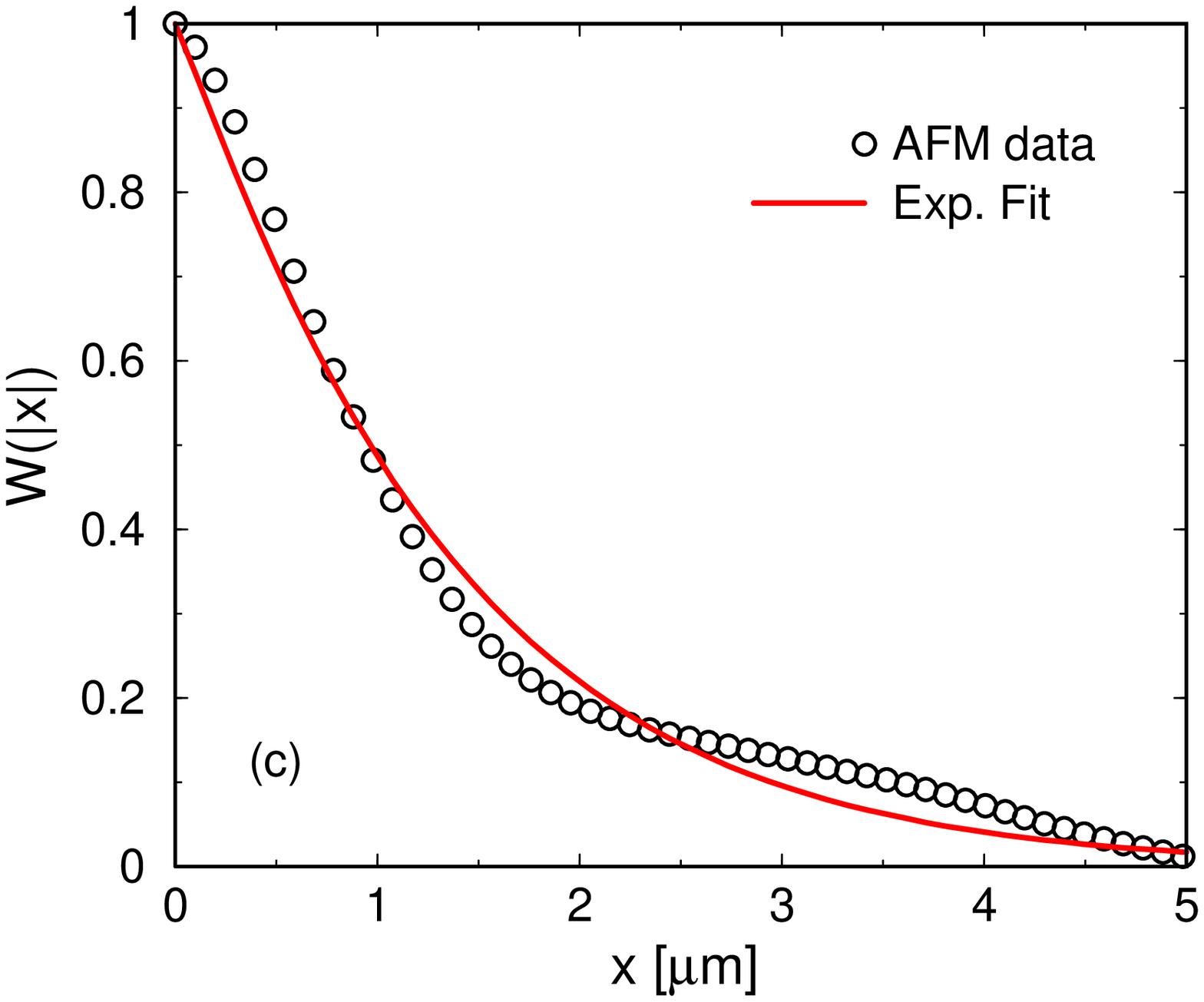}
      \caption{(Color online) Experimental height measurements
        obtained from Atomic Force Microscopy~(AFM) measurements on a
        linear low density polyethylene~(LLDPE) film.  The material
        was of the Narrow (molecular weight distribution) Metallocene
        grade. (a) A contour plot of the experimental height function
        measured over a $50\times50\, (\mu m)^2$ quadratic area.  The
        color code is so that red corresponds to $0.015\, \mu m$ and
        blue to $-0.015\, \mu m$. (b) The height distribution function
        $P(\zeta)$ calculated from the AFM data (open circles) and a
        Gaussian fit (solid line) corresponding to $\sigma=0.04\,\mu
        m$.  (c) The height-height correlation function $W(|x|)$
        obtained from the AFM data (open circles) and fitted with an
        exponential correlation function $W(|x|)=\exp(-|x|/a)$ (solid
        line) corresponding to a correlation length of $a=1.3\,\mu
        m$.}
    \label{fig:experimental-surf-data} 
\end{figure}

To investigate this further, we will consider a melt blown film of
linear low density polyethylene~(LLDPE). The film and LLDPE type was
referred to as {\em narrow molecular weight distribution metallocene},
and described in detail, in Ref.~\cite{4x}.
The haze of the film was measured by a spherical haze-meter (Diffusion
System, type M57) according to the standard~\cite{HazeStandard1}, and
(in transmission, at normal incidence) one obtained for this material
a haze of $18.9\%$.  Light scattering at the surface, as well as in
the bulk, contributed to this haze value.  The bulk contribution was
estimated by measuring the haze of films coated by glycerol on both
sides. Glycerol has a refractive index of $1.47$, which matches
closely the refractive index of the blown films (approximately $1.5$).
Under this condition the haze was significantly reduced (to $5\%$ for
the film mentioned above, leaving, crudely speaking, about $14\%$ haze
resulting from the surface roughness). The films that were
experimentally studied in Ref.~\cite{4x} (with that of
Fig.~\ref{fig:experimental-surf-data} being among them) gave a haze in
the range $12$--$30\%$. When embedded in glycerol, the corresponding
values were reduced to $3$--$6\%$. The film with the highest haze was
measured to be $31.9\%$, and reduced to about $1/10$ of this value
when submerged in glycerol ($3\%$). This indicates that the
contribution to haze from surface roughness dominates over the
contribution from the bulk. This is consistent with the assumption
made in the theory section of this paper (cf.
Sec.~\ref{Sec:Scatt-Theory}).

The surface topography of the above mentioned polymer film was
measured and served as the basis for the characterization put forward
in Fig.~\ref{fig:experimental-surf-data}.  The solid lines in
Figs.~\ref{fig:experimental-surf-data}(b) and (c) represent a Gaussian
(Fig.~\ref{fig:experimental-surf-data}(b)) and an exponential fit
(Fig.\ref{fig:experimental-surf-data}(c)) to the height distribution
and height-height correlation function respectively.  These functions
are characterized by a root-mean-square roughness of $\sigma=0.04\,\mu
m$ and a correlation length of $a=1.3\,\mu m$. Notice that the
functional fits performed in Figs.~\ref{fig:experimental-surf-data}(b)
and (c) are of reasonable quality and in particular for the height
distribution function. In the one-dimensional Monte Carlo simulation
results to be presented below, one has used the functions represented
by the solid lines in Figs.~\ref{fig:experimental-surf-data}(b) and
(c) as a basis for generating the underlying ensemble of surface
realizations.

\begin{figure}
  \centering
  \includegraphics*[width=12cm]{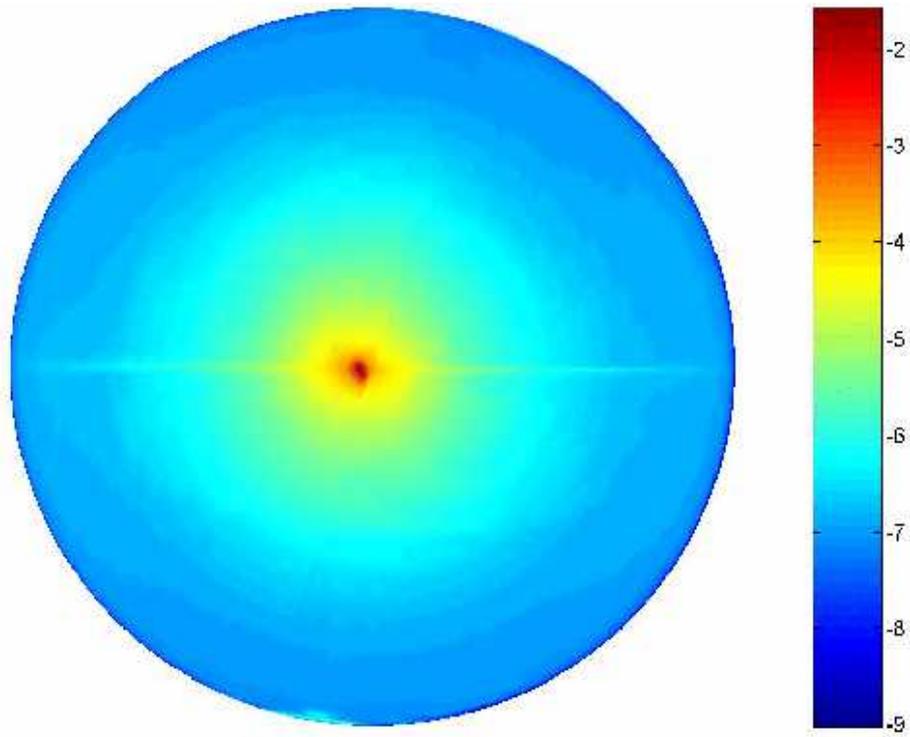} 
  \caption{ (Color online) Experimental results for the angular
    distribution of the light being transmitted through a melt blown
    LLDPE film of the type described and partly characterized in
    Fig.~\protect\ref{fig:experimental-surf-data}.  The (mean)
    thickness of the film was $d=50\mu m$, and the wavelength of the
    light being incident normally onto the top mean surface was
    $\lambda=0.6328\mu m$.  It is the logarithm of the transmitted
    intensity in arbitrary units that is presented. The measurements
    were conducted with a spectro-photo-goniometer built by SINTEF.
    The weak anisotropy seen in the transmitted intensity is caused by
    the polymers being preferentially oriented in the flow direction
    (vertical direction in the figure).}
   \label{fig:Measurements}
\end{figure}

In Fig.~\ref{fig:Measurements} the logarithm of the experimentally
obtained angular intensity distribution, $\log I(\theta,\phi)$, of
light being transmitted through the polymer film system described
above is depicted. In obtaining these results, a HeNe-laser at
wavelength $\lambda=0.6328\mu m$ was used as a source for the
unpolarized, normal incident light ($\theta_0=0^\circ$). As expected,
a strong specular transmission peak as well as a large dynamical range
in intensity (almost seven orders of magnitude) are observed. 
Moreover, a weak anisotropy in the angular distribution of the
transmitted light can be observed in Fig.~\ref{fig:Measurements} as
represented by the horizontal line of enhanced intensity.  This
anisotropy is caused by the polymers being partially oriented along
the direction of the flow in the production of the polymer film. 
In Fig.~\ref{fig:Measurements} this direction corresponds to the
vertical.  Except from the weak anisotropy the angular distribution of
the transmitted light is rather isotropic. 

\begin{figure}
  \centering
  \includegraphics*[width=10cm]{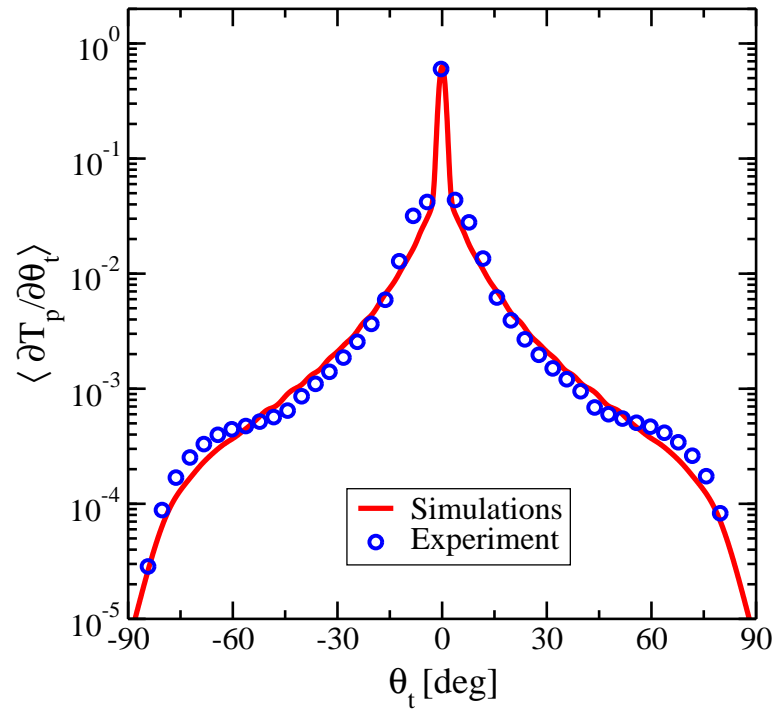}
  \caption{(Color online) The mean differential transmission
    coefficient $\left<\partial T_p/\partial\theta_t\right>$ {\it vs.}
    angle of transmission $\theta_t$ for light of wavelength
    $\lambda=0.6328\, \mu$m impinging normally ( $\theta_0=0^\circ$)
    onto a $d=40\mu m$ thick LLDPE film ($\varepsilon_1=2.25$). The
    experimental results~(open circles) corresponds to a cut through
    Fig.~\protect\ref{fig:Measurements}. Due to the use of arbitrary
    units in the experiment, the amplitude of the measurements was
    adjusted to fit that of the simulation results.  In obtaining the
    simulation results~(solid line), the surface parameters
    represented by the solid lines in
    Figs.~\protect\ref{fig:experimental-surf-data} were used, {\it
      i.e.}  the parameters used were $\sigma=0.04 \mu m$ and
    $a=1.3\mu m$ for the exponentially correlated rough surface. In
    order to replicate the unpolarized incident light used in the
    experiment, the simulation results were averaged over the $s$- and
    $p$-polarized results.  The thickness of the film was also here
    $d=40\mu m$.  }
    \label{fig:Comparison} 
\end{figure}

In Fig.~\ref{fig:Comparison}, a horizontal cut through the center of
Fig.~\ref{fig:Measurements} is represented by open symbols. The solid
line in this same figure represents the rigorous (one-dimensional)
Monte Carlo simulation results for the mean differential transmission
coefficient.  In obtaining this latter result a film geometry of mean
thickness $d=40\mu m$ was used where the uncorrelated upper and lower
rough (one-dimensional) interfaces were described by the parameters
that were derived from the measured surface topography of the film
(see Figs.~\ref{fig:experimental-surf-data}). In particular, these
surfaces were characterize by the functions represented by the solid
lines of Figs.~\ref{fig:experimental-surf-data}(b) and (c) (see also
the caption of Fig.~\ref{fig:Comparison} for the surface parameters).
From Fig.~\ref{fig:Comparison} it is observed that there is a quite
reasonable agreement between the measured and simulated angular
distribution of the transmitted light. It is at the largest angles of
transmission that the discrepancy starts to emerge, while the angular
distribution in the central part, that represents the main part of the
transmitted energy, seems to be well accounted for by the simulation
result.  For the largest transmission angles the Monte Carlo result
seems to underestimate the transmitted power. This situation is in
fact not unexpected; In the experimental measurements scattering from
the bulk is also present, while such effects has not been taken into
consideration in the approach used to produce the solid line of
Fig.~\ref{fig:Comparison}. It is, in fact, well known that bulk
scattering tends to enhance the transmitted power into large angles of
transmission~\cite{bulk}.

As mentioned above, one by direct measurements found the haze in
transmission for normal incidence to be ${\cal H}_{2D}(0^\circ)\approx
0.189$ for the experimental sample (with both surface and bulk
randomness). From the Monte Carlo simulation results one on the other
had found a haze of ${\cal H}_{1D}(0^\circ)\approx 0.054$. However, in
order to be able to compare this result with the experimentally
available value~(${\cal H}_{2D}(0^\circ)$), one has to introduce a
(multiplicative) constant that accounts for the difference between the
one- and two-dimensional geometry (assuming normal incident
light). Doing so results in a haze of about $14$--$15\%$, a result
that compares favorably to the experimental haze value resulting from
surface randomness.
Hence, one has demonstrated that the one dimensional Monte Carlo
approach can be applied to predict reasonably well the haze of an
isotropic experimental sample at normal incidence. This is indeed
encouraging results, but further work is needed, however, to determine
the full region of applicability of the Monte Carlo approach.

\section {Conclusions}
\label{Sec:Conclusions}    

In conclusion, we have studied the dependence of haze and gloss with
the parameters that normally are used to characterize randomly rough
surfaces --- the rms-roughness $\sigma$ and the height-height
correlation length $a$. Based on phase-perturbation theory, we have
derived analytic expressions that represent approximations to haze and
gloss for a one-dimensional Gaussian rough surface. It is demonstrated
that haze(gloss) increases(decreases) with $\sigma/\lambda$ as
$\exp(-A(\sigma/\lambda)^2)$ and decreases(increases) with $a/\lambda$
in a way that depends on the specific form of the correlation function
being considered. This latter dependence enters into the expression
for haze and gloss as the average of the power spectrum taken over a
small interval, $\Delta q$, around zero momentum transfer.  In the
limit $\Delta qa\ll 1$ one obtains that the haze and gloss depend
linearly on the correlation length. 

The range of validity for these approximate expressions to haze and
gloss put forward in this paper were assessed by rigorous numerical
Monte Carlo simulations. They were found to agree remarkably well over
large regions of parameter space.  Furthermore, some experimental
results for the angular distribution of the light being transmitted
through a polymer film was presented. It was found to fit reasonably
well with the prediction from the Monte Carlo simulations, and
consequently the predicted value of haze was found to agree relatively
well.

\section* {Acknowledgments}
It is a pleasure to acknowledge Rudie Spooren at SINTEF for having
conducted the transmitted intensity measurements presented in
Fig.~\ref{fig:Measurements}.  This work was in part financed by the
Borealis Group and the Research Council of Norwegian.


\appendix

\section{Phase perturbation theory}
\label{App}
 
In this appendix we will describe the so-called phase perturbation
theory, as well as deriving some analytical expressions, on which the
main text relays. What has become known as the phase-perturbation
theory, was originally developed by J.\ Shen and A.\ A.\
Maradudin~\cite{Shen} for non-penetrable media. The method was later
extended to one-dimensional randomly rough penetrable media by
S\'anchez-Gil {\it et al.}~\cite{SG}. The method has also been
formulated in an explicit reciprocal way~\cite{Rosa}.  In
phase-perturbation theory it is the phase of the field that is
determined perturbatively~\cite{Tsang-v3}, and it has proven well
suited for reflectivity studies~\cite{SG,Chaikina}.  One of the
interesting features of this pertubative method is that in the large
limit of $a/\lambda$, with $a$ being the surface correlation length
and $\lambda$ the wavelength of the incident light, it reduces to the
more well-known Kirchhoff
approximation~\cite{Beckmann,BassFuks,Ogilvy,Tsang-v1} for a
non-penetrable medium. Furthermore, as $a$ becomes comparable to
$\lambda$, phase-perturbation theory represents a correction to the
Kirchhoff result. An additional practical advantage of
phase-perturbation theory is that it remains its analytic form also
for penetrable or absorbing media, something that is not the case for
the Kirchhoff approximation.
 
\begin{figure}
  \centering
  \includegraphics[width=10cm]{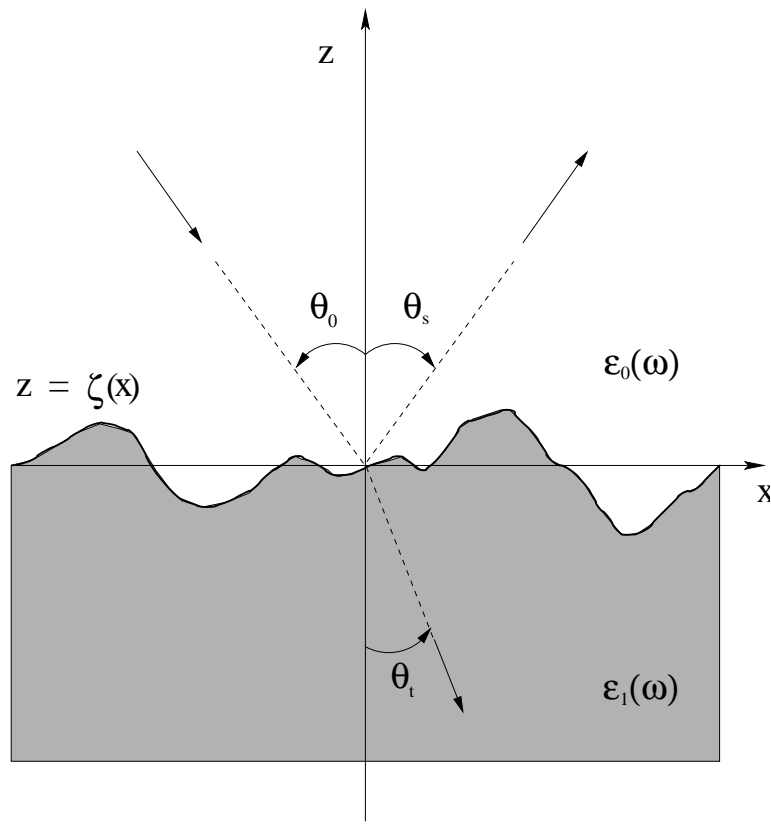}
\caption{The scattering geometry used in this
  study. The rough surface is defined by $z=\zeta(x)$. The region
  above the surface, $z>\zeta(x)$, is assumed to be vacuum
  ($\varepsilon_0(\omega)=1)$, while the medium below is a dielectric
  characterized by a frequency-dependent dielectric function
  $\varepsilon_1(\omega)$. Notice for which direction the angle of
  incident ($\theta_0$), scattering ($\theta_s$), and transmission
  ($\theta_t$) are defined as being positive. An angle of transmission
  is only well-defined if the lower medium is transparent, {\it i.e.} 
  if $Re\, \varepsilon_1(\omega)>0$. }
    \label{fig:Geometry} 
\end{figure}

Let us start our discussion of phase-perturbation theory and what can
be derived from it, by letting $U(q|k)$ collectively denote either the
reflection or transmission amplitudes, $R_\nu(q|k)$ or $T_\nu(q|k)$,
introduced in Sec.~\ref{Sec:MDRC}. By taking into account the
boundary condition at the rough surface, and assuming that there is no
down-going (up-going) scattered (transmitted) waves even close to the
rough interface~\footnote{In the language of rough surface scattering
  this is called the Rayleigh hypothesis~\cite{Rayleigh,Tsang-v1} in
  honor of Lord Rayleigh who first suggested its use. Formally this
  hypothesis amounts to assuming that the asymptotic expressions for
  the field above and below the surface,
  Eqs.~(\ref{eq:field-expansion}), can be used all the way down to the
  rough interface, and consequently used to fulfill the boundary
  conditions.} a single integral equation for $U(q|k)$ can be
derived~\cite{Rayleigh,ThesisIngve} (the reduced Rayleigh equation).
Based on this integral equation one can derive the following
expression for amplitude
\begin{subequations} 
\begin{eqnarray}
  \label{eq:amplitude}
  U(q|k) &=& u_0(k) \int^\infty_{-\infty} dx \; e^{-i(q-k)x} e^{i\Lambda(q|k)\zeta(x)}
\end{eqnarray}
where $u_0(k)$ is the Fresnel coefficient for the corresponding planar
geometry. In writing this expression we have introduced 
\begin{eqnarray}
  \label{app:Lambda}
  \Lambda(q|k) &=& \left\{
    \begin{array}{cll}
      \alpha_0(q)+\alpha_0(k), & \qquad & \mbox{in reflection} \\
        \alpha_1(q)-\alpha_0(k), & \qquad & \mbox{in transmission}
    \end{array}
  \right. . 
\end{eqnarray}
\end{subequations} 
It should be noticed that if we were dealing with a more complicated
scattering geometry then the one depicted in Fig.~\ref{fig:Geometry},
say a film geometry, the amplitude could still be expression in the
form (\ref{eq:amplitude}) if only one of the interfaces were randomly
rough. In this case $\Lambda(q|k)$ will take on another form then the
one given above. 

In order to make contact with observable we will need an expression
for $|U(q|k)|^2$. In fact, more precisely it is the average over this
quantity that we should be interested in since the surface is randomly
rough. With Eq.~(\ref{eq:amplitude}), we find
\begin{eqnarray}
  \label{eq:amp-avr}
  \left<\left| U(q|k) \right|^2\right>
            &=&
      \left|u_0(k)\right|^2 \int^\infty_{-\infty} dx
            \int^\infty_{-\infty} dx'  \;
          e^{-i(q-k)(x-x')} \left< e^{i\Lambda(q|k)[\zeta(x)-\zeta(x')]} \right>,
\end{eqnarray}
where the average, denoted by $\left<\cdots\right>$, is assumed to be
taken over an ensemble of surface realizations of the surface profile
function $\zeta(x)$. Furthermore, we have here assumed that
$\Lambda(q|k)$ is real (or close to being real), as it will be
automatically in reflection. 

With the change of variable $x'=x+u$ the above equation becomes 
\begin{eqnarray}
  \label{eq:amp-avr-1}
  \left<\left| U(q|k) \right|^2\right>
            &=&
      \left|u_0(k)\right|^2 \int^\infty_{-\infty} dx
            \int^\infty_{-\infty} du \;
          e^{i(q-k)u} \left< e^{i\Lambda(q|k)\Delta \zeta(u)} \right>,
\end{eqnarray}
with $\Delta\zeta(u)=\zeta(x)-\zeta(x+u)$.  If $\zeta(x)$ is a {\em
  stationary} random process, then the average in
Eq.~(\ref{eq:amp-avr-1}) will be independent of $x$.  Therefore, the
$x$-integration in the same equation will give the contribution $L$,
that is the length along the $x$-direction of the rough surface. 
Furthermore, if in addition to being stationary $\zeta$ also is a
Gaussian random process, as we assume here according to
Sec.~\ref{Sec:Geometry}, then $\Delta \zeta(u)$ will also be a
Gaussian random variable. Hence, the average in
Eq.~(\ref{eq:amp-avr-1}) can be done analytically with the result that
\begin{subequations}
  \label{EQ:Amp}
\begin{eqnarray}
  \label{eq:amp-avr-2}
  \left<\left| U(q|k) \right|^2\right>
            &=&
      L \left|u_0(k)\right|^2 \int^\infty_{-\infty} du \;
          e^{i(q-k)u} e^{-\sigma^2\Lambda^2(q|k)[1-W(u)] }
              \nonumber   \\
       &=&  \left|u_0(k)\right|^2  J(q|k),
\end{eqnarray}
where
\begin{eqnarray}
  \label{eq:app:J-ing}
  J(q|k) &=&  L\, e^{-\sigma^2\Lambda^2(q|k) }
          \int^\infty_{-\infty} du \;
          e^{i(q-k)u} e^{\sigma^2\Lambda^2(q|k)W(u) },
\end{eqnarray}
\end{subequations}
with $W(u)$ being the surface correlation function as defined in
Sec.~\ref{Sec:Geometry}. 

With Eqs.~(\ref{EQ:Amp}) and (\ref{eq:mean-DRC-DTC}), one for the
mean DRC or DTC, collectively denoted $\left<\partial
  U/\partial\theta\right>$, finally obtains
\begin{eqnarray}
  \label{eq:APP:MDRC-MDTC}
  \left<\frac{\partial U}{\partial\theta}\right> &=&
    \frac{1}{L}\frac{\varepsilon_m}{\sqrt{\varepsilon_0}}
    \frac{\omega}{2\pi c} \frac{\cos^2\theta}{\cos\theta_0}
    \left| u_0(k) \right|^2 J(q|k). 
\end{eqnarray}



%



\newpage

\newpage

\begin{thebibliography}{99}

\bibitem{1x} 
  J. L. White, Y. Matsukura, H. J. Kang, and H. Yamane,
  Inter. Polym. Process. {\bf 1}, 83 (1987). 

\bibitem{2x} 
  M. S. Edmondson and S. E. Pirtle, J. Plast. Film Sheeting {\bf 9}, 334 (1993). 

\bibitem{3x} 
  M. B. Johnson, G. L. Wilkes, A. M. Sukhadia, and D. C. Rohlfing,
  J. Appl. Polym. Sci. {\bf 77}, 2845 (2000). 

\bibitem{4x}	
  E. Andreassen, \AA. Larsen, K. Nord-Varhaug, M. Skar,
  H. {\O}ys{\ae}d, Pol. Eng. Sci. {\bf 42}, 1082 (2002). 

\bibitem{5x} 
  M. S. Pucci and R. N. Shroff, Polym. Eng. Sci. {\bf 26}, 569 (1986). 

\bibitem{6x} 
  A. M. Sukhadia, J. Plast. Film Sheeting {\bf 16}, 54 (2000). 

\bibitem{7x} 
  H. Zweifel, {\sl Plastics Additives Handbook}, 5th ed., (Hanser, Munich, 2001). 

\bibitem{8x} 
  S. Cheruvu, F. Y.-K. Lo, S. C. Ong, and T.-K. Su, International patent WO 95/13317 (1995).  

\bibitem{9x} 
  H. Ashizawa, J. E. Spruiell, and J. L. White, Polym. Eng. Sci. {\bf 24}, 1035 (1984). 

\bibitem{10x} 
  P. F. Smith, I. Chun, G. Liu, D. Dimitriev, J. Rasburn, and
  G. J. Vancso, Polym. Eng. Sci. {\bf 36}, 2129 (1996). 

\bibitem{11x} 
  F. C. Stehling, C. S. Speed, and L. Westerman, Macromolecules {\bf 14}, 698 (1981). 

\bibitem{12x} 
  A. Larena, and G. Pinto, Polym. Eng. Sci. {\bf 33}, 742 (1993). 

\bibitem{13x} 
  S. W. Shang and R. D. Kamla, J. Plast. Film and Sheeting {\bf 11}, 21 (1995). 

\bibitem{14x} 
  F. M. Willmouth, {\sl In Optical properties of polymers}, p.265,
  G. H. Meeten (Ed.), (Elsevier, London, 1986). 

\bibitem{HazeStandard1}
  ASTM Standard D 1003-95: {\sl Standard Test Method for Haze and
    Luminous Transmittance of Transparent Plastics}, 2000.
  
\bibitem{HazeStandard2}
 ASTM Standard E 430-91, {\sl Standard Test Method for Measurement
 of Gloss of High-Gloss Surfaces by Goniophotometry}, 1991.


\bibitem{15x}	
  Y. Kikuta, Y. Miyasaka, T. Asuke, Bunseki Kagaku {\bf 45}, 347 (1996). 

\bibitem{16x}	
  N.D. Huck, P.L. Clegg, SPE Trans. {\bf 1}, 121 (1961). 

\bibitem{17x} 
  M. Kojima, J. H. Magill, J. S. Lin, and S. N. Magonov,
  Chem. Mater. {\bf 9}, 1145 (1997). 

\bibitem{18x} 
  L. Wang, T. Huang, M. R. Kamal, A. D. Rey, and J. Teh,
  Polym. Eng. Sci. {\bf 40}, 747 (2000).   

\bibitem{19x} 
L. Wang, M. R. Kamal, and A. D. Rey, Polym. Eng. Sci. {\bf 41}, 358 (2001).   

\bibitem{20x}	
  A. M. Sukhadia , D. C. Rohlfing, M. B. Johnson, G. L. Wilkes,
  J. Appl. Polym. Sci. {\bf 85}, 2396 (2002). 

\bibitem{Rayleigh} J.\ W.\ S.\ Rayleigh, {\sl The Theory of Sound,
    Vols. 1 and 2} (Dover Publications, New York 1945, originally
  published 1876). 

\bibitem{Beckmann} P.\ Beckmann and A.\ Spizzichino, {\sl The
scattering from electromagnetic waves from rough surfaces} (Artech
House, 1963). 

\bibitem{BassFuks}
  F.G. Bass and I.M. Fuks, {\sl Wave scattering from statistically
    rough surfaces} (Pergamon Press, Oxford, UK, 1979). 

\bibitem{Ogilvy} 
  J.\ A.\ Ogilvy, {\sl Theory of wave scattering from
    random rough surfaces} (IOP Pub., Bristol, 1991). 
  

\bibitem{Tsang-v1} L.\ Tsang, J.A.\ Kong, and K.-H. Ding, {\sl
    Scattering of electromagnetic waves: Theories and Applications},
  vol. 1  (John Wiley \& Sons, Inc.,
  2000)

\bibitem{Tsang-v2} L.\ Tsang, J.A.\ Kong, K.-H. Ding and C.O.\ Ao,  {\sl Scattering
    of electromagnetic waves: Numerical Simulations}, vol. 2 (John
  Wiley \& Sons, Inc., 2001)

\bibitem{Tsang-v3} L.\ Tsang and  J.A.\ Kong, {\sl Scattering
    of electromagnetic waves: Advanced topics}, vol. 3 (John
  Wiley \& Sons, Inc. 2001)

\bibitem{Born} M.\ Born and E.\ Wolf, {\sl Principles of optics}, 7th
(expanded) edition  (Cambridge Univ. Press, Cambridge, 1999). 


\bibitem{ThesisIngve}
   I.\ Simonsen, Ph.D. thesis (The Norwegian University of
  Science and Technology, Trondheim, 2000).\\
  Available from  http://www.phys.ntnu.no/\~{}ingves/Thesis/

\bibitem{Ingve-condmat}
   I.\ Simonsen, {\sl A random walk through rough surface scattering
   phenomena}, arXiv:cond-mat/0408017, 2004.

\bibitem{Maradudin-Review}
  A.V. Zayats, I.I. Smolyaninov, and A.A. Maradudin,
  Phys. Rep. {\bf 408}, 131 (2005)

\bibitem{SA1}
 I. Simonsen, D. Vandembroucq, and S. Roux, 
 Phys. Rev. E {\bf 61}, 5914 (2000).
 

\bibitem{SA2}
 I. Simonsen, D. Vandembroucq, and S. Roux, 
 J. Opt. Soc. Am. A {\bf 18}, 1101 (2001). 




\bibitem{PhysRep1995} 
  X.-L.\ Zhou and S.-H.\ Chen, Phys.\ Rep.\ {\bf 257}, 223 (1995).


\bibitem{Sinha1988}
 S.K.\ Sinha, E.B.\ Sirota, S.\ Garoff, and H.B.  Stanley,
 Phys.\ Rev.\ B {\bf 38}, 2297 (1988).   


\bibitem{deBoer}
 D.K.G.\ de Boer, Phys.\ Rev.\ B {\bf 49}, 5817 (1994),
 {\it ibid.} Phys.\ Rev.\ B {\bf 51}, 5297 (1995).


\bibitem{Andreev}
  A.V.\ Andreev, Phys.\ Lett.\ A {\bf 219}, 349 (1996).

\bibitem{Alex}
  A.A.\ Maradudin and T.A.\ Leskova, Waves Random Media {\bf 7}, 395  (1997).




\bibitem{Designer1}
 A.A. Maradudin, I. Simonsen, T.A. Leskova, and E. R. M\'endez, 
 Opt. Lett. {\bf 24}, 1257 (1999). 


\bibitem{Designer2}
  E.R. M\'endez {\it et al.}, Appl. Phys. Lett.  {\bf 81}, 798 (2002).

\bibitem{24x} 
  J.C. Stover, {\sl Optical scattering -- Measurements and Analysis}, 2nd
  ed. (SPIE Optical Engineering Press, Bellingham, WA, 1995). 

\bibitem{25x} 
  E. C. Teague, Annals CIRP {\bf 30}, 563 (1981). 


\bibitem{Barrera} 
  R.\ Alexander-Katz and R.\ G.\ Barrera, 
  J. Polym. Sci. B: Polym. Phys. {\bf 36}, 1321 (1998). 


\bibitem{27x} 
  E. Bahar, B. S. Lee, G. R. Huang, and R. D. Kubik, Radio Science
  {\bf 30}, 545 (1995). 

\bibitem{28x} 
  W. T. Welford, Optical and Quantum Electronics {\bf 9}, 269 (1977). 
  

\bibitem{Jackson}
 J.\ D.\ Jackson, {\sl Classical Electrodynamics}, 2nd
edition  (John Wiley \& Sons, New York, 1975). 


\bibitem{Gloss}
 M.E. Nadal and E.A. Thompson, 
Journal of Coatings Technology {\bf 72} (911), 61 (2000). 


\bibitem{Mendez}
  R.\ R.\ M\'endez, R.\ G.\ Barrera, and R.\ Alexander-Katz,
 Physica A {\bf 207}, 137 (1994). 


\bibitem{Billmeyer} 
 F.\ W.\ Billmeyer and Y.\ Chen, Color Research and
  Application {\bf 10}, 219 (1985). 
  

\bibitem{DesginerSurfaces}
  A.\ A.\ Maradudin, I.\ Simonsen, T.\ A.\ Leskova, 
        and E.\ R.\ M\'endez,
        Opt. Lett. {\bf 24},1257 (1999). 

\bibitem{LambertianDiff}
   A.\ A.\ Maradudin, I.\ Simonsen, T.\ A.\ Leskova, 
        and E.\ R.\ M\'endez,
    Waves Random Media {\bf 11}, 529 (2001). 

\bibitem{Shen}
  J.\ Shen and A.\ A.\ Maradudin,
   Phys.\ Rev.\ B {\bf 22}, 4234 (1980). 


\bibitem{Rosa}
R. M. Fitzgerald and A.A. Maradudin,
Waves Random Media {\bf 4}, 275 (1994). 


\bibitem{SPP}
  V.\ M.\ Agranovich, D.\ L.\ Mills (editors),  {\sl Surface
  Polaritons},  (North-Holland, Amsterdam, 1982). 
 

\bibitem{Stegun}
    M.\ Abramowitz and I.\ A.\ Stegun, {\em Handbook of Mathematical
    Functions}, (Dover, New York, 1964).


\bibitem{SG}
  J.\ A.\ S\'anchez-Gil, A.\ A.\ Maradudin, and E.\ R.\ M\'endez,
  J. Opt. Soc. Am. A {\bf 12}, 1547 (1995). 

\bibitem{our-gloss-paper}
  I.\ Simonsen, {\AA}.G.\ Larsen, E.\ Andreassen, E.\ Ommundsen, and
  K.\ Nord-Varhaug,
  Phys.\ Stat.\ Sol.\ (b) {\bf 242}, 2995 (2005).  

\bibitem{AnnPhys}
 A.\ A.\  Maradudin, T.\ Michel, A.\ R.\ McGurn, and E.\ R.\ M\'endez,
    Ann. Phys. {\bf 203}, 255 (1990). 

\bibitem{Wavelength_dependence} 
  I. Simonsen, T.A. Leskova, and A.A. Maradudin, O. Hunderi,
      Proc. Int. Soc. Opt. Eng. {\bf 4100}, 65 (2000).


\bibitem{bulk}
  D.\ Bicout and C.\ Brosseau,
 J. Phys. I France {\bf 2}, 2047 (1992).

 


\bibitem{Chaikina}
  E.\ I.\ Chaikina {\it et al.}, Appl. Opt. {\bf 37}, 1110 (1998). 





\end{thebibliography}
\end{document}